\documentclass[oldversion]{aa}

\usepackage{txfonts}
\usepackage{graphicx}
\usepackage{color}
\usepackage{natbib}
\usepackage{amssymb}

\newcommand{\dd}{\mathrm{d}}

\newcommand{\mbh}{\ensuremath{M_\bullet}\,}
\newcommand{\mbu}{\ensuremath{M_\mathrm{Bulge}}\,}

\newcommand{\mbhbu}{\ensuremath{M_\bullet}-\ensuremath{M_\mathrm{Bulge}\,}}
\newcommand{\mbhsig}{\ensuremath{M_\bullet}-\ensuremath{\sigma_\ast\,}}
\newcommand{\ler}{\ensuremath{\log \lambda\,}}

\begin{document}
\title{Selection effects in the black hole-bulge relation and its evolution
       }

\author{A. Schulze \and L. Wisotzki}
\institute{Leibniz-Institut f\"ur Astrophysik Potsdam (AIP), An der Sternwarte 16, 14482 Potsdam, Germany }
\offprints{A. Schulze, \email{aschulze@aip.de} }

\date{Received / Accepted }
\abstract{We present an investigation of sample selection effects that influence the observed black hole - bulge relations and its evolution with redshift. We provide a common framework in which all kinds of selection effects on the $M_\bullet-$bulge relations can be investigated, but our main emphasis is on the consequences of using broad-line AGN with virial estimates of black hole masses and their host galaxies to search for evolution in the BH - bulge relation.  We identified relevant sources of bias that were not discussed in the literature so far. A particularly important effect is caused by the fact that the active fraction among SMBHs varies considerably with BH mass, in the sense that high-mass BHs are less likely to be active than lower mass ones. In the connection with intrinsic scatter of the BH - bulge relation this effect implies a bias towards a low BH mass at given bulge property. This effect adds to the bias caused by working with luminosity or flux limited samples that were already discussed by others. A quantitative prediction of these biases requires (i) a realistic model of the sample selection function, and (ii) knowledge of relevant underlying distribution functions: The distribution of spheroid properties such as velocity dispersions or masses; the active fraction as a function of BH mass, or alternatively the active black hole mass function; and the Eddington ratio distribution function. We employed our formalism together with recently determined distribution functions to investigate how much existing studies of the $M_\bullet$-bulge relation using AGN hosts might suffer from selection biases. For low-redshift AGN samples we can naturally reproduce the flattening of the relation observed in some studies without having to invoke intrinsic differences in the BH - bulge relation between active and inactive galaxies. When extending our analysis to higher redshift samples we are clearly hampered by limited empirical constraints on the various relevant distribution functions. Using a best-guess approach for these distributions we estimate the expected magnitude of sample selection biases for a number of recent observational attempts to study the $M_\bullet$-bulge evolution. In no case do we find statistically significant evidence for an evolving $M_\bullet$-bulge relation. While the observed apparent offsets in \mbh/\mbu\ from the local relation can be quite large, the sample selection bias estimated from our formalism is typically of the same magnitude. We suggest a possible practical approach to circumvent several of the most problematic issues connected with AGN selection; this could become a powerful diagnostic in future investigations.
}

\titlerunning{Selection effects on the black hole-bulge relations and its evolution}
\authorrunning{Schulze \& Wisotzki}
\maketitle

\section{Introduction}  \label{sec:bias_intro}
Supermassive black holes appear to be ubiquitous in the centers of massive galaxies \citep{Kormendy:1995}. Furthermore, they show tight correlations with the properties of the spheroidal components of their host galaxies, e.g. with the stellar velocity dispersions \citep{Ferrarese:2000,Gebhardt:2000, Tremaine:2002,Gultekin:2009}, bulge luminosities and masses \citep{Magorrian:1998,Marconi:2003,Haering:2004,Sani:2010}, or concentration indices \citep{Graham:2001}. These relations have been established by direct dynamical measurement of the black hole mass in a few dozens of nearby galaxies, mainly using stellar dynamics \citep[e.g.][]{vanderMarel:1998,Emsellem:1999,Gebhardt:2003} or gas dynamics \citep[e.g.][]{Ferrarese:1996,Marconi:2001,DallaBonta:2009}. 

Nevertheless, significant uncertainties in these relations persist. The high mass and the low mass regimes are still poorly constrained. There may also be systematic effects on the determined masses; for example, not accounting for triaxiality of the galaxy \citep{vandenBosch:2010} or not including a dark matter halo in the dynamical models \citep{Gebhardt:2009,Schulze:2011} can bias the dynamical measurement. Also the intrinsic scatter in these relations is still not well established \citep{Gultekin:2009}. Furthermore, there is evidence that late-type galaxies and in particular pseudobulges do not follow the same relations as early-type galaxies \citep{Hu:2008,Graham:2008,Greene:2010,Kormendy:2011,Jiang:2011}.

The black hole - bulge relations contain important information, as they imply a connection between the growth of a black hole and the evolution of its host galaxy. A common picture invokes AGN feedback to shut down star formation and self-regulate black hole accretion. Models building on this scenario are able to reproduce the local black hole - bulge relations in numerical simulations \citep{DiMatteo:2005,Sijacki:2007,Booth:2009} and semi-analytically \citep{Kauffmann:2000,Cattaneo:2005,Croton:2006b,Bower:2006,Somerville:2008,Marulli:2008}. A correlation between black hole mass and host galaxy mass will also be tightened or may even be generated naturally within a merger driven galaxy evolution framework  \citep{Peng:2007, Hirschmann:2010, Jahnke:2010}.

Essential constraints on the origin of the black hole - bulge relation can be inferred from their redshift evolution. There are several theoretical predictions, based on numerical simulations \citep{Robertson:2006,Hopkins:2007b,DiMatteo:2008,Johansson:2009,Booth:2010,Dubois:2011} as well as on semi-analytic models \citep{Wyithe:2003,Croton:2006,Malbon:2007,Hopkins:2009,Lamastra:2010}. Although the details are still far from being settled, they tend to predict an increase in the $M_\bullet/M_\mathrm{Bulge}$ ratio with redshift, while only a weak or even negative evolution  in the \mbhsig relation is expected.

These models need to be confronted with observations. Several different approaches have been followed in the last years to  observationally constrain the evolution in the black hole - bulge relations. These range from more indirect arguments, to direct estimates of $M_\bullet$ and the respective bulge property. Constraints on integrated quantities can be gained from the black hole mass function \citep{Merloni:2004b,Hopkins:2006,Shankar:2009,Somerville:2009,Kisaka:2010,Willott:2010}. \citet{Bluck:2011} studied X-ray selected AGN, employing Eddington ratio arguments, to constrain black hole - bulge coevolution.

Direct dynamical determinations of $M_\bullet$ are not feasible outside of the very local universe. Therefore, the most commonly adopted approach resorts to broad line AGN, for which black hole masses are easily accessible through the so-called 'virial estimator' \citep[e.g.,][]{McLure:2002,Vestergaard:2006}. The main challenge for the determination of the $M_\bullet$-bulge relationship from AGN samples is the determination of the bulge properties, hampered by the bright nuclear point source of the AGN, which may well outshine the entire galaxy.

Evolution in the \mbhsig relation has been studied either by measuring stellar velocity dispersions directly \citep{Woo:2006,Woo:2008,ShenJ:2008} or by using the widths of narrow emission lines as surrogates for $\sigma_\ast$ \citep{Shields:2003,Shields:2006,Salviander:2007}. The \mbhbu relation has been investigated from QSO host galaxy luminosities \citep{Peng:2006a,Peng:2006b,Treu:2007,McLeod:2009,Decarli:2010,Bennert:2010,Targett:2011}, by estimating the stellar masses utilising colour information \citep{Schramm:2008,Jahnke:2009,Merloni:2010}, or by dynamical mass measurements \citep{Inskip:2011}. Also obscured AGN with detectable broad lines have been used \citep{Sarria:2010,Nesvadba:2011}, for which the determination of stellar masses is less problematic. \citet{McLure:2006} inferred the \mbhbu ratio up to $z=2$ by matching distributions of radio-loud QSOs and radio galaxies. At the highest redshifts, CO rotation curves have been used to determine dynamical masses for a few individual objects \citep{Walter:2004,Riechers:2008,Riechers:2009,Wang:2010}. All these methods have their own advantages and drawbacks. Nevertheless, there is increasing evidence for apparent evolution, in a sense of high-$z$ SMBHs being more massive at given bulge mass than in the local universe. 

However, it has been suspected that these results may be biased due to sample selection effects. One such bias could arise from the fact that luminous AGN on average tend to have massive black holes. The intrinsic scatter in the BH - bulge relation then generates a bias towards a higher $M_\bullet/M_\mathrm{Bulge}$ ratio. Such a bias has been pointed out by several authors \citep{Adelberger:2005,Fine:2006,Salviander:2007}, but received major attention by the work of \citet{Lauer:2007} who discussed it in some detail \citep[but see also ][]{Peng:2010}. A further effect recently discussed by \citet{Shen:2010} is caused by the uncertainty in virial black hole mass estimates in connection with the steep decrease of the active black hole mass function.

In this paper we argue that besides the above-mentioned effects there are even more potential biases that need to be considered. While previous authors mostly focused on one particular source of bias, we aim at providing a comprehensive study of selection effects on observations of the \mbhbu relation and its evolution. We also investigate how well these biases can be quantified and corrected for. The paper is organised as follows: In section~\ref{sec:loc} we present our general framework to investigate the consequences of selection effects and show a first application to the quiescent black hole sample. Section~\ref{sec:agn} discusses several selection effects for AGN samples and their ramifications for the black hole-bulge relations. In section~\ref{sec:evo} we take into account redshift evolution effects. We discuss the implications of our results on observational studies in section~\ref{sec:bias_discus}. We finally conclude in section~\ref{sec:bias_conclu}. For the cosmological parameters we assume $H_0 = 70$~km~s$^{-1}$~Mpc$^{-1}$, $\Omega_\mathrm{m} = 0.3$ and $\Omega_\Lambda = 0.7$.

\section{The local \mbhbu relation} \label{sec:loc}

\subsection{General framework} \label{sec:loc-gen}
We start with some basic thoughts on the BH - bulge relation and its sensitivity to selection effects, postponing specific aspects of AGN samples to the next section. We adopt the following convention: Black hole masses are given by $\mu=\log M_\bullet$, while any of the various relevant bulge properties is denoted as $s$, with $s=\log \sigma_\ast$ or $s=\log M_\mathrm{Bulge}$, respectively. In some examples we focus on the \mbhbu relation, but the arguments equally hold for the \mbhsig relation. We explicitly do not imply one relation to be more fundamental than the other. 

The distribution of objects in the $\mu$-$s$ diagram is given by the bivariate distribution function of bulge property and black hole masses $\Psi(s,\mu)$. Thus $\Psi(s,\mu)\,\dd \mu \,\dd s$ gives the number of objects per Mpc$^3$ with galaxy property between $s$ and $s+\dd s$ and black hole mass between $\mu$ and $\mu+\dd \mu$. We assume the following parameterisation for most of the paper,
\begin{equation}
 \Psi(s,\mu)=g(\mu \, |\, s)\, \Phi_s(s) \ ,  \label{eq:psi}
\end{equation} 
where $\Phi_s(s)$ is the spheroid distribution function, e.g. the spheroid mass function, and $g(\mu \, |\, s)$ gives the probability of finding the black hole mass $\mu$, given $s$. If $s$ and $\mu$ are correlated, as suggested by the observations, then $g(\mu \, |\, s)$ corresponds to this correlation. In the following we  assume a linear relation $\mu= a + b s$ with log-normal intrinsic scatter $\sigma$, i.e.
\begin{equation}
 g(\mu \, |\, s) = \frac{1}{\sqrt{2 \pi} \sigma} \exp \left\lbrace - \frac{(\mu-a-bs)^2}{2 \sigma^2} \right\rbrace  \ . \label{eq:gms}
\end{equation} 
This parameterisation for the bivariate distribution function is not the only one possible. We comment on the observational justification of this form in the Appendix. While our quantitative results depend on the adopted parameterisation, most of our qualitative results are independent of the specific choice of $\Psi(s,\mu)$.

The corresponding distribution functions for $s$ and $\mu$ are given by marginalisation over the other variable,
\begin{equation}
 \Phi_s(s)= \int \Psi(s,\mu)\, \dd\mu = \Phi_s(s) \ , \label{eq:phis}
\end{equation} 
\begin{equation}
 \Phi_\bullet(\mu)= \int \Psi(s,\mu)\, \dd s = \int g(\mu \, |\, s)\, \Phi_s(s)\, \dd s \ .\label{eq:qbhmf}
\end{equation}
The integration over $\mu$ simply returns the spheroid mass function. The integration over $s$ provides the quiescent black hole mass function (BHMF), equivalent to the common approach for its determination \citep[e.g.][]{Yu:2004,Marconi:2004,Merloni:2004}. The first equality in Equations~(\ref{eq:phis}) and~(\ref{eq:qbhmf}) is independent of the special choice of $\Psi(s,\mu)$. In particular, as the galaxy distribution function is an observable, Equation~(\ref{eq:phis}) sets a constraint on the bivariate distribution $\Psi(s,\mu)$. Note that the specific shape of $\Psi(s,\mu)$ has direct consequences for the quiescent black hole mass function.

\begin{figure}
\centering
\resizebox{\hsize}{!}{\includegraphics[clip]{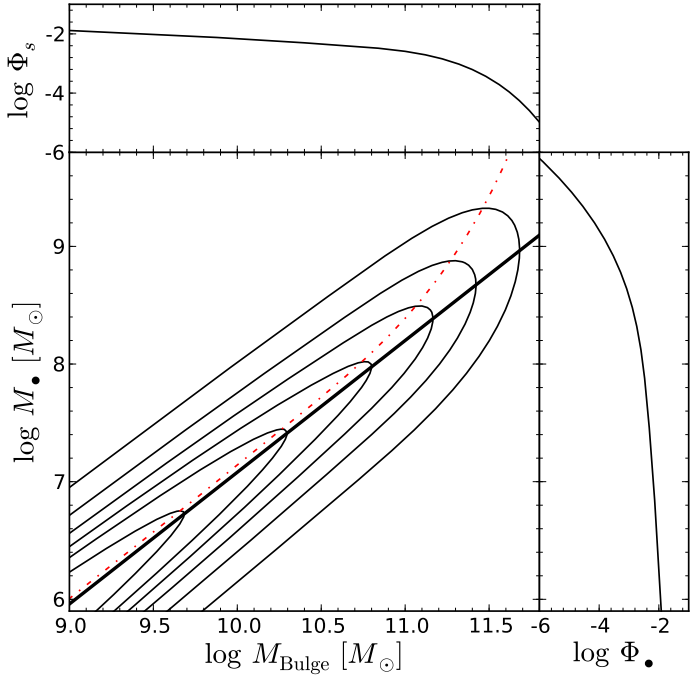}}
\caption{Bivariate probability distribution function of bulge masses and black hole masses with no selection effects. The contours indicate decreasing probability in logarithmic units. The thick black line shows the input \mbhbu relation from \cite{Haering:2004}. The red dot-dashed line shows the mean galaxy property at a given black hole mass. The upper panel shows the projection of the bivariate distribution function to the bulge mass, i.e. the spheroid mass function. The right panel shows the projection to black hole mass, i.e the quiescent black hole mass function.}
\label{fig:bhbu_dfs}
\end{figure}

Fig.~\ref{fig:bhbu_dfs} shows $\Psi(s,\mu)$, $\Phi_s(s)$ and $\Phi_\bullet(\mu)$ for the local universe, employing our parameterisation. We estimated the spheroid mass function from the early type and late type stellar mass functions of \citet{Bell:2003}. We computed the spheroid mass function as sum of the early type mass function and the mass function of the bulge components of late type galaxies. Since the bulge fraction in disk galaxies is still somewhat uncertain, we simply assumed an average value of $B/T=0.3$ to convert the late type mass function into a bulge mass function, consistent with current observations \citep[e.g.][]{Graham:2008b,Gadotti:2009}. This value also produces a black hole mass function that is consistent with determinations from the \mbhsig and $M_\bullet$-$L$ relations \citep{Marconi:2004}. For the \mbhbu relation we used the relation by \citet{Haering:2004} and an intrinsic scatter of $\sigma=0.3$~dex. For most of this paper we keep this value for the intrinsic scatter fixed; we comment on the consequences of changing it when appropriate.
 
The distribution shown in Fig.~\ref{fig:bhbu_dfs} would be obtained for a strictly volume limited sample of galaxies with $M_\mathrm{Bulge}$ and \mbh measurements. In practice, the distribution will be affected by the way that the sample was constructed. This is formally accounted for by a selection function $\Omega$, defined as the probability of observing an object of a given bulge mass, black hole mass and further possible selection criteria, such as for instance redshift or AGN luminosity. The selection function can thus be written as $\Omega(s,\mu,\theta)$, where $\theta$ refers to the set of additional parameters present as selection criteria. The observed apparent bivariate distribution is then given by
\begin{equation}
  \Psi_\mathrm{o}(s,\mu)= \int \Omega(s,\mu,\mathbf{\theta})\, \Psi(s,\mu)\, p_\theta(\theta)\,\dd \theta \ , \label{eq:psio}
\end{equation} 
where $p_\theta(\theta)$ is a set of normalised distribution functions of the parameters $\theta$.
Depending on $\Omega$, the observed \mbhbu relation can be significantly biased if the selection effects are not taken into account. However, this requires a proper knowledge of the selection function.

The bivariate distribution $\Psi_\mathrm{o}(s,\mu)$, when normalised to one, represents the full probability distribution of the expected $M_\bullet-M_\mathrm{Bulge}$ relation. However, it may happen that we are not so much interested in the full distribution, but rather in the mean relation. This is given by the mean black hole mass at a given bulge property
\begin{equation}
 \langle \mu \rangle (s) = \frac{\int \mu\, \Psi_\mathrm{o}(s,\mu)\, \dd \mu}{\int \Psi_\mathrm{o}(s,\mu)\, \dd \mu} \ .\label{eq:meanm}
\end{equation} 
If no selection effects are present then $\langle \mu \rangle (s)$ will be identical to the input $M_\bullet$-bulge relation, $\langle \mu \rangle = a + b s$. Alternatively we can consider the mean bulge property at a given black hole mass 
\begin{equation}
 \langle s \rangle (\mu) = \frac{\int s \Psi_\mathrm{o}(s,\mu)\, \dd s}{\int \Psi_\mathrm{o}(s,\mu)\, \dd s} \ .\label{eq:means}
\end{equation}
Even without any selection effects, $\langle s \rangle (\mu)$ does not simply correspond to the inverse relation. It deviates from it quite strongly at the high mass end where the space density in the galaxy distribution function is decreasing. This effect is illustrated as the red dot-dashed line in Fig.~\ref{fig:bhbu_dfs} and was already discussed by \citet{Lauer:2007}. It is not an observational bias but just a direct consequence of the different projections chosen, combined with the steep decrease of the galaxy distribution function. In general, under the presence of selection effects $\langle s \rangle (\mu)$ may change as well.

In observational studies of the evolution of the $M_\bullet$-bulge relation, the sample is often compared to the local relation in terms of a single value, the offset from the local relation. The sample bias for this offset is given by
\begin{equation}
 \langle \Delta \mu \rangle = \frac{\iint (\mu-a-b s) \, \Psi_\mathrm{o}(s,\mu)\, \dd \mu \,\dd s}{\iint \Psi_\mathrm{o}(s,\mu)\, \dd \mu \,\dd s} \ . \label{eq:meandm}
\end{equation} 
Alternatively, the mean offset of the galaxy property from the local relation is given by
\begin{equation}
\langle \Delta s \rangle = \frac{\iint (s-\alpha-\beta \mu) \, \Psi_\mathrm{o}(s,\mu)\, \dd \mu \,\dd s}{\iint \Psi_\mathrm{o}(s,\mu)\, \dd \mu \,\dd s} = -\frac{1}{b} \langle \Delta \mu \rangle \ , \label{eq:meands}
\end{equation} 
where we use the inverse relation $s=\alpha+\beta \mu$ with $\alpha=-a/b$ and $\beta=1/b$. This directly leads to $\langle \Delta s \rangle=-1/b\,\langle \Delta \mu \rangle$, without any specific assumption on $\Psi_\mathrm{o}(s,\mu)$. Thus the biases in $\langle \Delta \mu \rangle$ and $\langle \Delta s \rangle$ should always be directly proportional to each other.

\begin{figure*}
\centering
\includegraphics[width=8cm,clip]{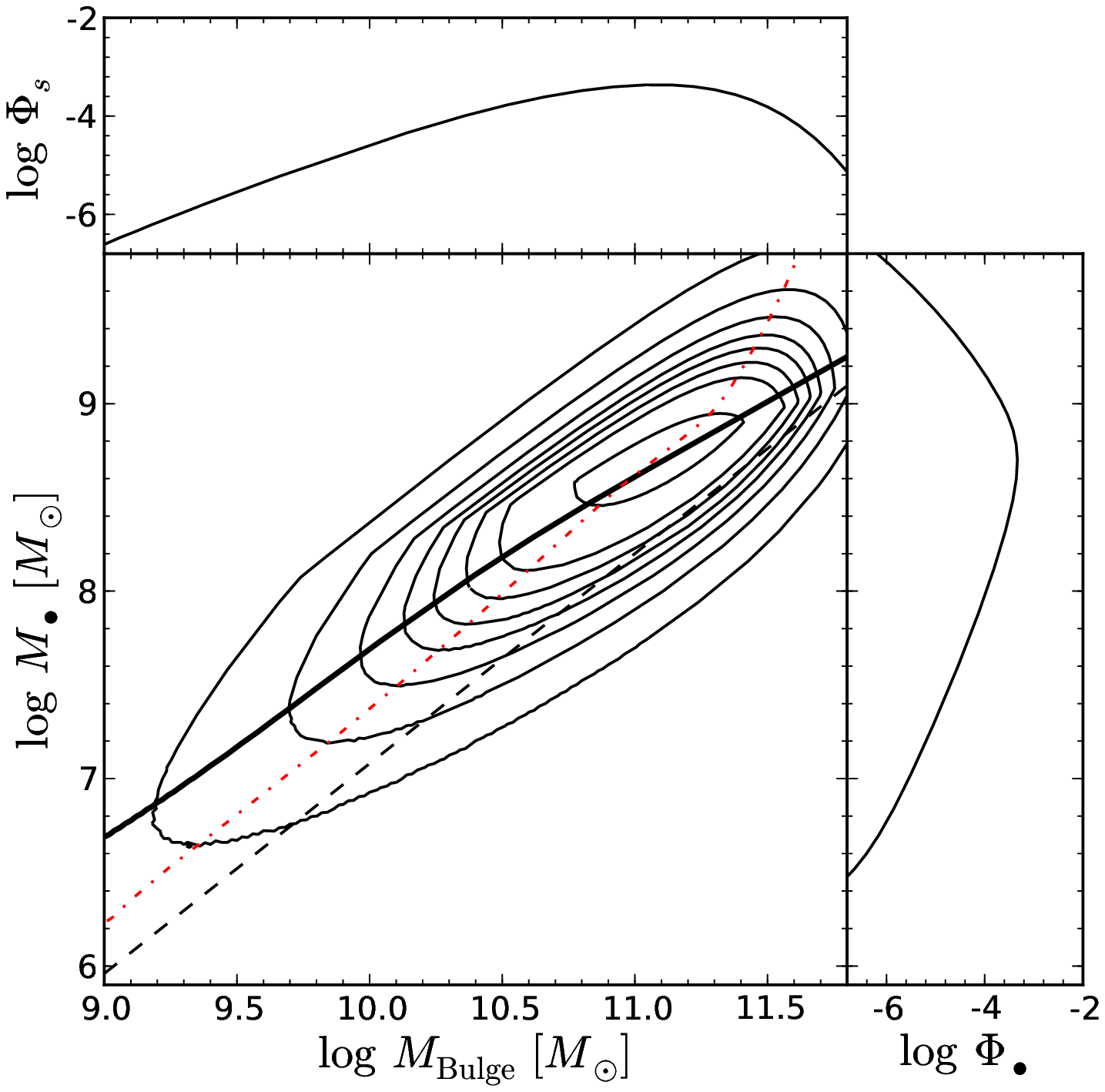} \hspace{1cm}
\includegraphics[width=8cm,clip]{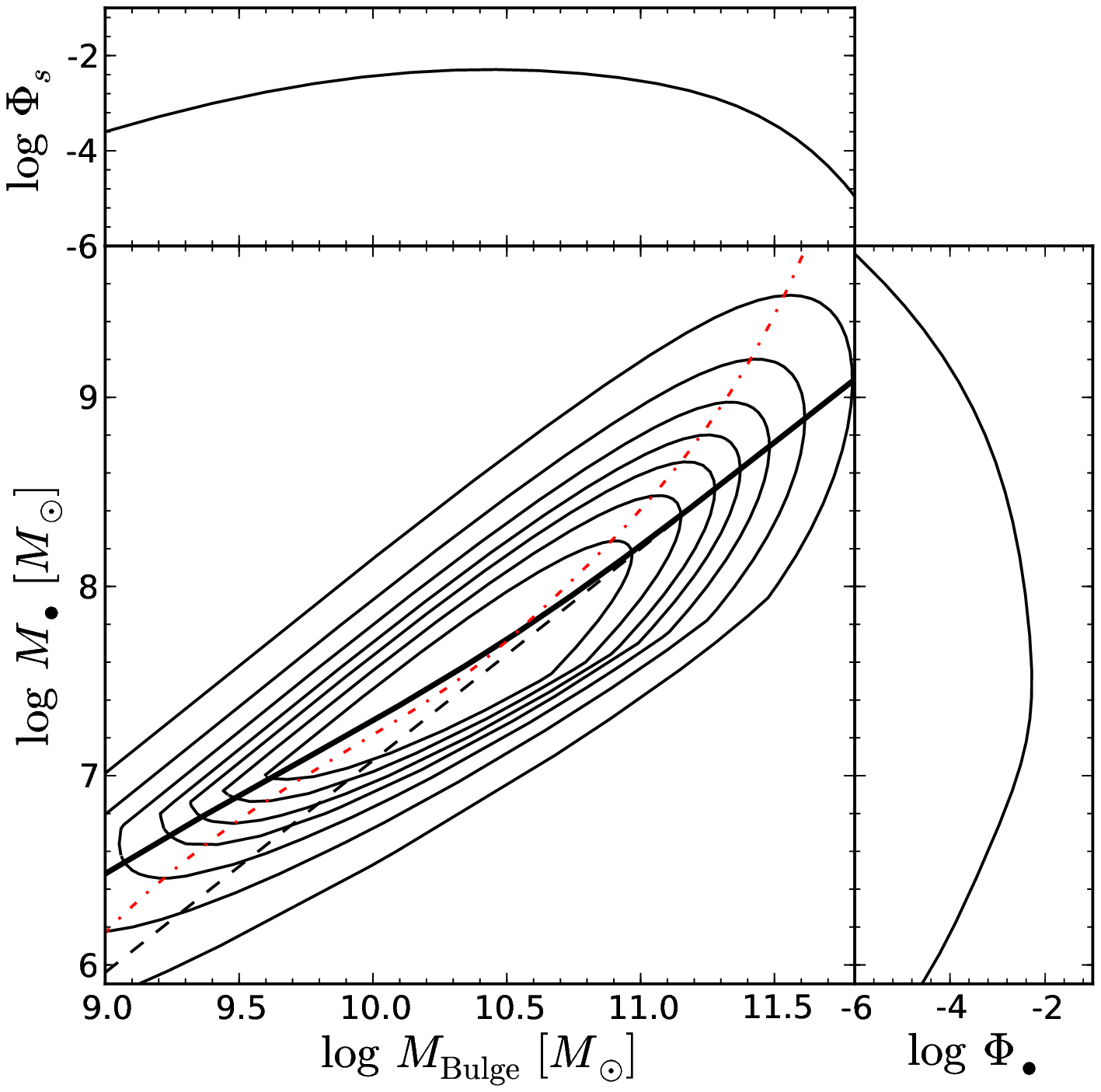}
\caption{The bivariate probability distribution function and its projections for a sample of black holes with dynamically measured masses, under the presence of a sample selection depending on the resolution of the black hole's sphere of influence. A threshold of $R_\mathrm{inf}/d_\mathrm{res}>1$ (left panel) or $R_\mathrm{inf}/d_\mathrm{res}>0.1$ (right panel) is applied to the sample. The dashed black line shows the input \mbhbu relation, the thick black solid line highlights the 'observed' relation $ \langle \mu \rangle (s)$, and the red dashed dotted line indicates the 'observed' relation $\langle  s \rangle (\mu)$}.
\label{fig:bhbu_rinf}
\end{figure*}

Already the well-studied samples of a few dozen local galaxies with reasonably well-measured dynamically-based black hole masses will invariably be affected by selection effects. However, these samples are very inhomogeneous, and presumably they cannot be described by a single well-defined selection function, which hampers a proper investigation of potential selection effects. \citet{Yu:2002} and \citet{Bernardi:2007} reported on a possible bias in the dynamical black hole mass sample identified through finding a discrepancy in the $\sigma_\ast-L$ relation compared to the SDSS. More recently, \citet{Gultekin:2009} discussed a bias induced by culling the sample based on resolving the spheres of influence of the black holes. Before discussing selection effects on AGN samples in detail, we investigate this bias to illustrate the general applicability of Equation~(\ref{eq:psio}).

\subsection{Bias by sphere of influence resolution} \label{sec:loc-dyn}
Dynamical black hole mass determinations need to spatially resolve the region of gravitational influence of the central black hole on the stellar velocity distribution. The size of this region is commonly estimated by the 'sphere of influence' of radius $R_\mathrm{inf}=GM_\bullet \sigma_\ast^{-2}$. It has been argued that dynamical black hole masses are unreliable for $R_\mathrm{inf}/d_\mathrm{res}<1$ and therefore should be excluded from the sample \citep[e.g.][]{Ferrarese:2005}. Here $d_\mathrm{res}$ is the spatial resolution of the observations. \citet{Gultekin:2009} demonstrated that rejecting galaxies below a line $M_\bullet \propto \sigma_\ast^2$ leads to a bias in the $M_\bullet$-bulge relations. \citet{Batcheldor:2010} pointed out that it would even be possible to artificially generate a \mbhsig correlation via this selection effect, while \citet{Gultekin:2011} excluded this possibility at least for the most extreme case. \citet{Gultekin:2009} argued against culling the sample based on the sphere of influence to avoid this bias. However, even without any active rejection the same effect may be effectively present, either due to an implicit target selection or through the mere ability to detect a black hole. The condition $R_\mathrm{inf}/d_\mathrm{res}\geq1$ is no strict limit for the reliability of a black hole detection, but it is clear that for $R_\mathrm{inf}/d_\mathrm{res}\ll1$ no black hole can be detected dynamically. 

We now illustrate the results of this bias. We assume that we can detect all black holes above a certain threshold in resolution of the sphere of influence, $R_\mathrm{inf}/d_\mathrm{res}>r_\mathrm{min}$. We keep the angular resolution of our hypothetical survey fixed at $0.1\arcsec$. The selection function is then given by
\begin{equation}
\Omega(s,\mu,d) = \left\{ \begin{array}{rl}
                 \displaystyle 1 & \quad \mathrm{for} \quad R_\mathrm{inf}/d_\mathrm{res}>r_\mathrm{min} \\
                  0 & \quad \mathrm{else} 
                \end{array} \right.
\end{equation} 
where $d$ is the distance of the galaxy. The observed bivariate probability distribution follows as 
\begin{equation}
  \Psi_\mathrm{o}(s,\mu)=\frac{3}{d_2^3-d_1^3} \, \int_{d_1}^{d_2} \Omega(s,\mu,d)\, \Psi(s,\mu)\, d^2\,\dd d \, .\label{eq:psirinf}
\end{equation} 
For the purpose of illustration we again use the \mbhbu diagram. We convert bulge masses into velocity dispersions assuming a fixed mass-to-light ratio for the $r$-band and using the $\sigma_\ast-L_r$ relation from \citet{Bernardi:2007b}. We cover the distance range from $1$ to $30$~Mpc. In Fig.~\ref{fig:bhbu_rinf} we show the \mbhbu diagram for two thresholds in the sphere of influence resolution, $r_\mathrm{min}=1$ and $r_\mathrm{min}=0.1$. We additionally plot the 'observed' mean \mbhbu relation $\langle \mu \rangle (s)$ for both cases as thick black solid line and the 'observed' mean relation $\langle s \rangle (\mu)$ as red dashed-dotted line. The bias induced by whether or not resolving the sphere of influence leads to a flatter slope, a higher normalisation and a smaller intrinsic scatter, as already shown by \citet{Gultekin:2009}. See also their work for an extensive Monte Carlo investigation of this bias. 

While a significant bias is induced only for a high threshold $r_\mathrm{min}$, the probability distribution is affected even for lower values, most severely at the low mass end. Also the relation  $\langle s \rangle (\mu)$ is biased in both cases. The selection along lines of $M_\bullet \propto \sigma_\ast^2$ is clearly visible in the probability contours. However, it is hard to quantify how much the actually observed samples of black holes are actually biased by this selection effect, due to their inhomogeneous character. But these calculations show that selection effects are a concern in general, also for dynamical black hole mass measurements.

\section{Biases of broad-line AGN samples} \label{sec:agn}
Selection effects are a major concern for samples selected as broad line AGN. For these objects the black hole mass can be easily estimated using the virial method. This makes them so valuable for studies of the evolution in the $M_\bullet$-bulge relations, as at high redshift black hole masses cannot be determined by means of direct dynamical observations. We first discuss generic issues that are also inherent in low-$z$ AGN samples. The application to higher redshifts will be covered in section~\ref{sec:evo}.

Samples of non-active galaxies are typically drawn from the galaxy luminosity function. In contrast, AGN samples are effectively drawn from the luminosity function of broad-line AGN. This can induce non-trivial selection effects on the sample, as already emphasized by \citet{Lauer:2007}. Firstly, there will be a luminosity bias. Flux limited AGN surveys will on average contain more luminous objects then an 'ideal' volume-limited sample. Thus, the AGN will on average host more massive black holes than in the volume-limited case. The intrinsic scatter in the black hole-bulge relations, together with the steep decrease of the spheroid distribution function, will then cause a Malmquist type bias towards more massive black holes at a given spheroid mass. The strength of the bias depends on the luminosity limit, the shape of the spheroid distribution function, the intrinsic scatter in the $M_\bullet$-bulge relation, and the distribution of Eddington ratios. We will further discuss and illustrate these details below.

A second effect, so far not discussed in the literature, can be called an 'active fraction bias'. For any sample of broad line AGN, their black holes are not randomly drawn from the entire black hole population (as described by the BH mass function), because they are by definition active. Only a minority of all black holes is in an active state, quantified by the active fraction (or equivalently by the AGN duty cycle). If this active fraction is a function of $M_\bullet$ then the intrinsic scatter in the \mbhbu relation will induce an additional bias, as will be shown below. If, for example, the active fraction decreases with increasing black hole mass, then for a given spheroid mass it will be more probable to find a smaller mass black hole in an AGN sample, causing a bias towards a lower $M_\bullet/M_\mathrm{Bulge}$ ratio. Conversely, for an increasing active fraction we expect a positive bias, while for a constant ($M_\bullet$-independent) active fraction no such bias will occur. Thus, this bias can work in both directions, depending on the black hole mass dependence of the active fraction. It will add to the luminosity bias.

\begin{figure*}
\centering
\includegraphics[width=8cm,clip]{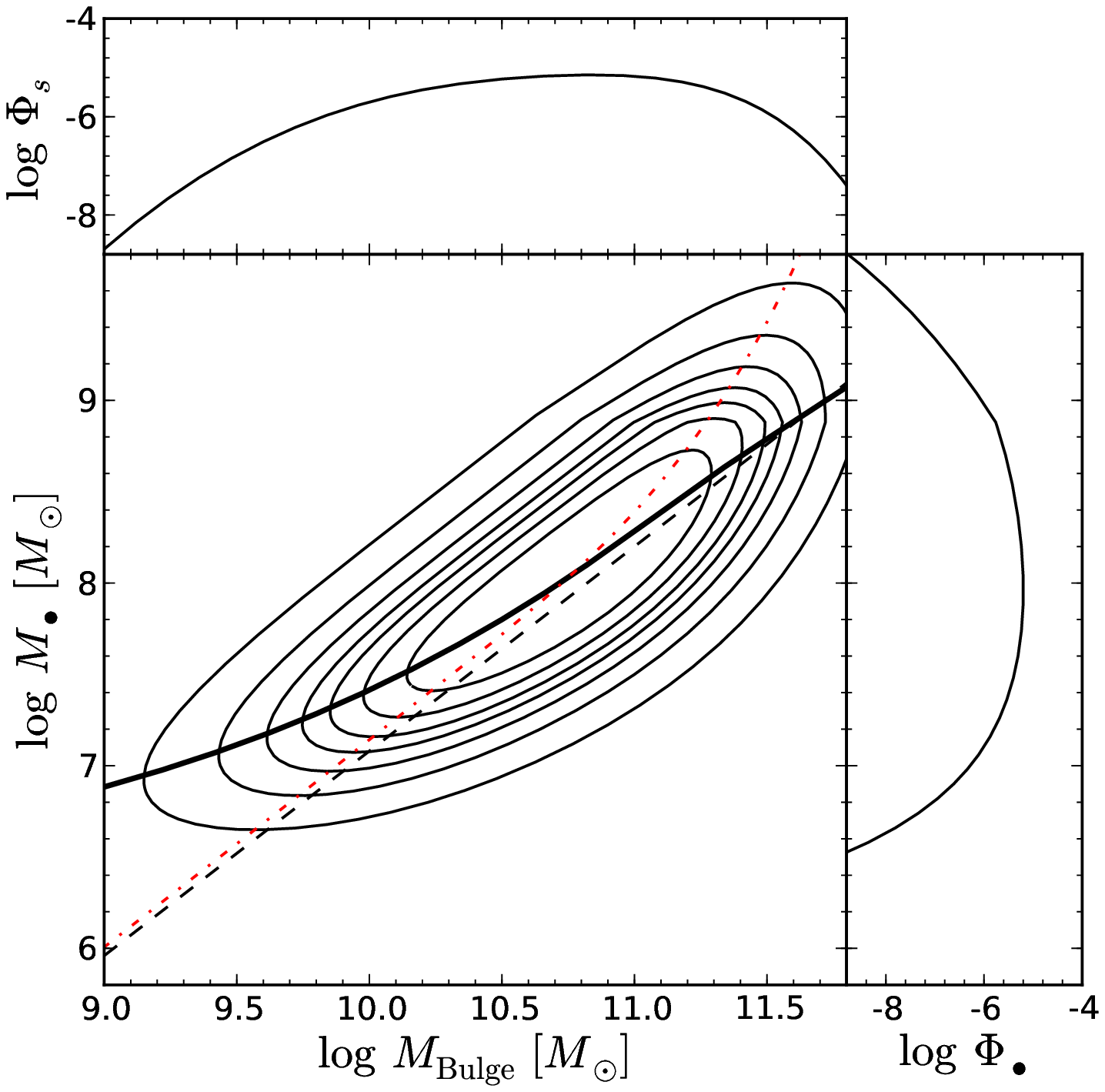} \hspace{1cm}
\includegraphics[width=8cm,clip]{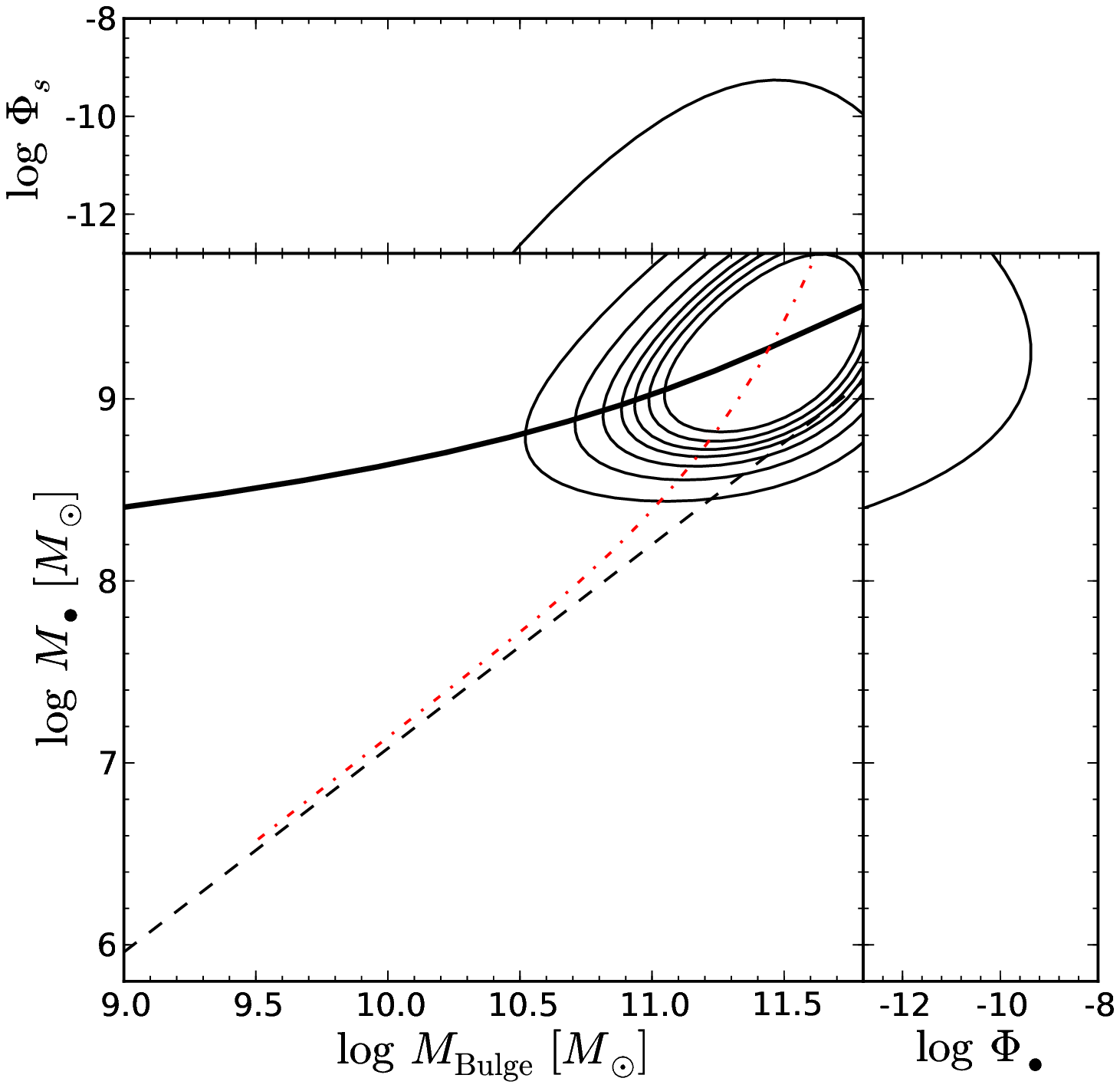}
\caption{Bivariate probability distribution function and its projections for a local type~1 AGN sample under the presence of a lower luminosity limit. The dashed black line again shows the input relation, the thick black solid line the observed relation $ \langle \mu \rangle (s)$, and the red dashed dotted line indicates the 'observed' relation $\langle  s \rangle (\mu)$ of this sample. A lower luminosity limit of $\log L_\mathrm{bol}>45$ (left panel) and $\log L_\mathrm{bol}>47$ (right panel) is applied, respectively.}
\label{fig:bhbu_agn}
\end{figure*}

\subsection{Luminosity limited samples} \label{sec:bias_llimit}
We now show analytically how these selection effects influence the $M_\bullet$-bulge diagram. Broad line AGN are drawn from the AGN luminosity function $\Phi_L(l)$, with $l\equiv\log L_\mathrm{bol}$. The luminosity of an AGN is produced by mass accretion onto a SMBH, and it is therefore determined by the black hole mass and the mass accretion rate. The latter is expressed in normalised units by the Eddington ratio $\lambda = L_\mathrm{bol}/L_\mathrm{Edd}$.
The Eddington luminosity is proportional to \mbh, thus the bolometric luminosity is given by $l = \log \lambda + \mu +38.1$. The AGN luminosity function $\Phi_L(l)$ (in logarithmic units) is then given by
\begin{equation}
\Phi_L(l)= \int p_\lambda(l-\mu)\, \Phi_{\bullet,\mathrm{a}}(\mu)\, \dd \mu \ . \label{eq:lf}
\end{equation} 
By $\Phi_{\bullet,\mathrm{a}}(\mu)$ we define the \textit{active} BHMF, where active black holes are defined as those that contribute to the corresponding AGN luminosity function. Thus, when restricting the AGN sample to type~1 AGN (showing broad lines), the active BHMF  includes only these type~1 AGN. The distribution function $p_\lambda(l-\mu)$ gives the probability of finding a black hole with mass $\mu$ given an AGN luminosity $l$, i.e. it corresponds to the normalised distribution function of Eddington ratios $\lambda$. We implicitly assume that the Eddington ratio distribution function is independent of black hole mass, i.e. $p_\lambda(\mu \, | \, l)=p_\lambda(l-\mu)=p_\lambda(\log \lambda)$. Type~1 AGN are observed to have Eddington ratios in the range $0.01<L_\mathrm{bol}/L_\mathrm{Edd}<1$. When applying a lower threshold to the Eddington ratio, not all black holes are currently in an active state. The active fraction is defined as the ratio between the active and the total black hole population, estimated as
\begin{equation}
p_\mathrm{ac}(\mu) = \Phi_{\bullet,\mathrm{a}}(\mu) / \Phi_{\bullet,\mathrm{q}}(\mu) \ ; \label{eq:ac}
\end{equation}  
with $\Phi_{\bullet,\mathrm{a}}(\mu)$ and $\Phi_{\bullet,\mathrm{q}}(\mu)$ as the active and quiescent BHMFs, respectively. Therefore, the AGN luminosity function can be expressed as 
\begin{equation}
\Phi_L(l)= \iint p_\lambda(l-\mu)\, p_\mathrm{ac}(\mu)\,g(\mu \, |\, s)\, \Phi_s(s) \, \dd s \dd \mu \ . \label{eq:lfo}
\end{equation}
There is thus a connection between the bulge property and the AGN luminosity, which is however smeared out by convolution with a set of additional distribution functions.

Following Equation~(\ref{eq:psio}), the bivariate black hole-bulge distribution function for a luminosity limited sample is given by 
\begin{equation}
\Psi_\mathrm{o}(s,\mu)= \int \Omega(s,\mu,l)\, p_\lambda(l-\mu)\,g(\mu \, |\, s)\, \Phi_s(s) \,\dd l \, .
\end{equation}
If we assume a fixed lower luminosity limit $l_\mathrm{lim}$, the selection function is
\begin{equation}
\Omega(s,\mu,l) = \left\{ \begin{array}{rl}
                 \displaystyle p_\mathrm{ac}(\mu) & \quad \mathrm{for} \quad l \geq l_\mathrm{lim} \\
                  0 & \quad \mathrm{else} 
                \end{array} \right.
\end{equation} 
The bivariate distribution function is then 
\begin{equation}
\Psi_\mathrm{o}(s,\mu)= p_\mathrm{ac}(\mu)\,g(\mu \, |\, s)\, \Phi_s(s) \, \int_{l_\mathrm{lim}}^\infty  p_\lambda(l-\mu)\,\dd l \, .\label{eq:psilum}
\end{equation}
Hence, $\Phi_L(l)$ and $\Psi_\mathrm{o}(s,\mu)$ are just different projections of a multivariate distribution function $\Psi_\mathrm{o}(s,\mu,l)$. The non-trivial difference between $\Psi(s,\mu)$ and $\Psi_\mathrm{o}(s,\mu)$ is the source of the bias. It is controlled by the three probability distributions, $p_\mathrm{ac}(\mu)$, $p_\lambda(l-\mu)$, and $g(\mu \, |\, s)$. The distribution of $p_\mathrm{ac}(\mu)$ regulates the active fraction bias, $p_\lambda(l-\mu)$ rules the luminosity bias and $g(\mu \, |\, s)$ controls the overall strength of the bias.

In Equation~(\ref{eq:psilum}) we can combine the parts governing the AGN selection by defining the selection function, integrated over AGN luminosity,
\begin{equation}
\Omega(\mu)= p_\mathrm{ac}(\mu)\, \int_{l_\mathrm{lim}}^\infty  p_\lambda(l-\mu)\,\dd l \, .\label{eq:omega_mu}
\end{equation}
and thus $\Psi_\mathrm{o}(s,\mu)=\Omega(\mu) \Psi(s,\mu)$.

The mean relations obtained from such a sample, affected by an AGN luminosity limit, are
\begin{equation}
 \langle \mu \rangle (s) = \frac{\int \mu\, \Psi_\mathrm{o}(s,\mu)\, \dd\mu}{\int \Psi_\mathrm{o}(s,\mu)\, \dd\mu} = \frac{\int \mu \Omega(\mu) g(\mu \, |\, s) \,\dd\mu}{\int \Omega(\mu) g(\mu \, |\, s)  \, \dd\mu} \ ; \label{eq:meanmagn}
\end{equation} 
\begin{equation}
 \langle s \rangle (\mu) = \frac{\int s \Psi_\mathrm{o}(s,\mu)\, \dd s}{\int \Psi_\mathrm{o}(s,\mu)\, \dd s} = \frac{\int s g(\mu \, |\, s) \Phi_s(s)\, \dd s}{\int g(\mu \, |\, s) \Phi_s(s) \,\dd s} \ . \label{eq:meansagn}
\end{equation} 
The relation $\langle \mu \rangle (s)$ is independent of the galaxy distribution function, but it is affected by the selection function. Thus, $\langle \mu \rangle (s) \neq a +bs$, as would be the case without selection effects. On the other hand $\langle s \rangle (\mu)$ is already intrinsically affected by the galaxy distribution function, but it is independent of the selection function. Therefore, $\langle s \rangle (\mu)$ contains information on the intrinsic relation, unaffected by the AGN luminosity bias and active fraction bias. This is an interesting fact, as it provides a potential route to study the evolution in the $M_\bullet$-bulge relations without having to account for AGN-specific selection effects. However, even if $\langle s \rangle (\mu)$ is not biased by the selection effects, the mean offset $\langle \Delta s \rangle$ is still biased, as shown by Equation~(\ref{eq:meands}), because we have to integrate over the entire observed black hole mass distribution.

We now first illustrate the AGN bias in the $M_\bullet$-bulge plane, and then investigate it in terms of a simple offset from the input relation. For this purpose, we need to know all the underlying distribution functions, for which we here adopt the $z\approx 0$ values. For $\Phi_s(s)$ we again use the above estimated spheroid mass function. For $g(\mu \, |\, s)$ we assume a log-normal distribution with intrinsic scatter $\sigma=0.3$~dex around the relation from \citet{Haering:2004}. These then define the quiescent BHMF via Equation~(\ref{eq:qbhmf}). If we further know the active BHMF, the active fraction is given by Equation~(\ref{eq:ac}). The active BHMF and the Eddington ratio distribution function (ERDF) $p_\lambda(l-\mu)$ for type~1 AGN in the local universe were determined by \citet[][hereafter SW10]{Schulze:2010}, and we use those values, adopting a modified Schechter function for the BHMF and a Schechter function for the ERDF. The SW10 active BHMF implies a significant decrease of the active fraction with increasing black hole mass, thus the active fraction bias will work towards a lower $M_\bullet/M_\mathrm{Bulge}$ ratio, opposite to the luminosity bias. Note that this parameterisation by construction ensures the consistency with the broad line AGN luminosity function via Equation~(\ref{eq:lfo}).

In Fig.~\ref{fig:bhbu_agn} we show the $M_\bullet$-$M_\mathrm{Bulge}$ diagram for a luminosity limited local AGN sample, for two lower luminosity limits, $l_\mathrm{min}=45$ and $l_\mathrm{min}=47$. At $z\sim0$ the latter case is rather unrealistic, as QSOs of these luminosities are extremely rare. However, as further discussed below, it is of importance for higher $z$ observations, where such luminosities are quite common in QSO samples. A luminosity limit of $l_\mathrm{min}=47$ roughly corresponds to the SDSS magnitude limit of $i\simeq19$ at $z\simeq3$ and to the SDSS high-$z$ limit of $i=20.2$ at $z\simeq6$.

For both luminosity limits the $M_\bullet$-$M_\mathrm{Bulge}$ relation is biased towards larger black hole masses at a given spheroid mass. This is already true for $l_\mathrm{min}=45$, but it is greatly enhanced for the high luminosity limit. The solid line again shows the mean black hole mass for a given spheroid mass (computed via Equation~(\ref{eq:meanm})), i.e. the apparent $M_\bullet$-$M_\mathrm{Bulge}$ relation obtained for the given sample. It is clearly visible that in general the bias is stronger for lower masses. Specifically, it is strongest close to the luminosity limit of the survey, which on average corresponds to the lowest mass black holes. The red dashed-dotted line again shows the mean spheroid mass for a given black hole mass (computed via Equation~(\ref{eq:means})). As noted, this relation is unaffected by the AGN selection and is identical to the one shown in Fig.~\ref{fig:bhbu_dfs}. However, it is not identical to the intrinsic relation, as it turns upwards at the high mass end due to the decrease in the spheroid mass function. Therefore, an integration over $\langle s \rangle (\mu)$ with a stronger weight on the high mass end will lead to a bias for the mean offset from the local relation, i.e for the total sample bias.

\begin{figure}
\centering
\resizebox{\hsize}{!}{\includegraphics[clip]{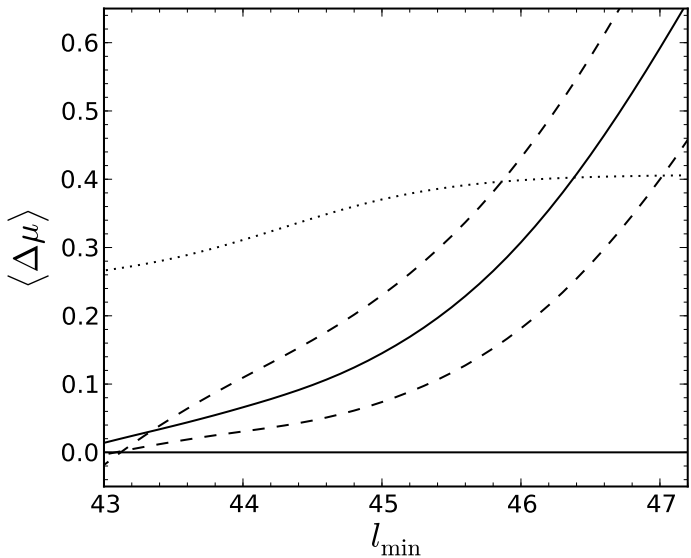}}
\caption{Predicted overall bias for a luminosity limited local ($z\approx 0$) type~1 AGN sample as a function of the low luminosity limit (in logarithmic units). The solid line shows our prediction for $\sigma=0.3$, while the dashed lines show the predictions for $\sigma=0.4$ (upper line) and $\sigma=0.2$ (lower line). The dotted line shows the prediction from \citet{Lauer:2007}, but for our local AGN luminosity function; see text for an explanation of the differences.}
\label{fig:dmact}
\end{figure}

The sample bias is greatly enhanced for a brighter luminosity limit. The steep decrease of the spheroid mass function and the intrinsic scatter in the relation itself induce an increase in the bias compared to the flat section of the spheroid mass function. This is shown in Fig.~\ref{fig:dmact}, where the mean offset from the input relation for the whole AGN sample, $\langle \Delta \mu \rangle $ (computed via Equation~(\ref{eq:meandm})), is plotted for a range of bolometric AGN luminosity limits, assuming the local distribution functions. Here we assumed our reference value for the intrinsic scatter, $\sigma=0.3$~dex, but we also indicate the dependence of the bias on the intrinsic scatter. A smaller intrinsic scatter also leads to a smaller sample bias, while a larger scatter causes a larger bias. By the dotted line we additionally show the bias predicted by \citet{Lauer:2007}, using their Equation~(25) and the local bolometric type~1 AGN luminosity function from \citet{Schulze:2009}. The main difference between their result and ours is that we not only use the AGN luminosity function, but evaluate the whole set of underlying distribution functions. In particular, we also take into account the active fraction bias that reduces the total bias, due to the decrease of the active fraction towards higher black hole masses. As a caveat we note that for very high luminosities, our knowledge of the underlying distributions (active and passive BHMFs, AGN luminosity function and \mbhbu relation) are observationally poorly determined.

\begin{figure}
\centering
\resizebox{\hsize}{!}{\includegraphics[clip]{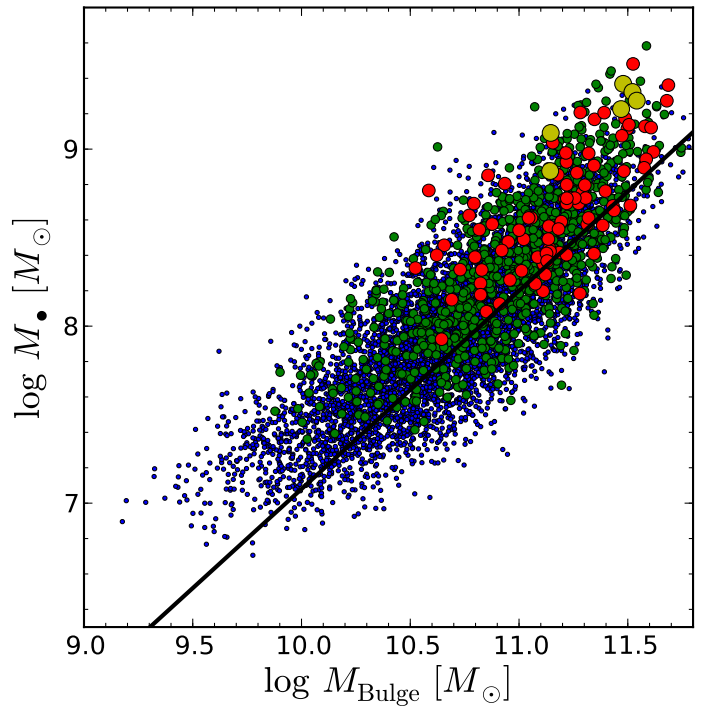}}
\caption{$M_\bullet$-$M_\mathrm{Bulge}$ diagram for a Monte Carlo simulation of a local AGN sample. The differently sized and coloured symbols correspond to different luminosity limits applied to the sample. Blue, green, red, and yellow symbols show the sample when culled at a bolometric luminosity of $l_\mathrm{min}=45$, $45.5$, $46$, and $46.5$ in logarithmic units. The solid black line shows the input relation for the sample from \citet{Haering:2004}.}
\label{fig:mclmin}
\end{figure}

We provide an additional illustration of these selection effects by Monte Carlo Simulations. We generated a large galaxy sample drawn from the spheroid mass function, and attributed black hole masses drawn from a log-normal distribution with mean from the \citet{Haering:2004} relation. We decided if the black hole is in an active state based on the probability $p_\mathrm{ac}(\mu)$. If it is active, an Eddington ratio was drawn from the SW10 Eddington ratio distribution function, which also sets the bolometric luminosity of the AGN. Now a number of specific selection criteria were applied to this sample, with the results presented in Fig.~\ref{fig:mclmin}. It is notable that for a bright luminosity limit most, if not all, AGN lie above the input $M_\bullet$-$M_\mathrm{Bulge}$ relation, while they are intrinsically drawn from it. The applied luminosity limit effectively corresponds to an unsharp lower limit on black hole mass. But at the same time also many massive black holes are excluded that accrete only at low rates. For low spheroid masses only positive outliers from the relation can be detected, while for increasing spheroid masses also black holes on and below the relation can still be detected. This causes the bias at low $M_\mathrm{Bulge}$.

In a realistic AGN sample there may be additional selection effects influencing the distribution of $M_\mathrm{Bulge}$. For example, usually a minimum host-to-nucleus ratio is required for the AGN host galaxy to be detected. This will correspond to some sort of cutoff in $M_\mathrm{Bulge}$, excluding low mass hosts. This additional selection effect will decrease the average bias for the sample, as the excluded objects are the most heavily biased. However, this will only be of importance if objects are actively eliminated during the sample construction based on these considerations.
Therefore, it is important to understand and quantify the selection effects as precisely as possible for a realistic prediction of the average bias of the given sample.

\subsection{Flux limited samples}
So far, we have only discussed purely luminosity limited samples. This approximation is valid for samples spanning small ranges in redshift. However, more generally we have to consider samples defined by limitations in apparent flux. (Note that luminosity limits may be present in addition, because the AGN need to be identified as such and separated from their host galaxies.) We first define the multivariate distribution function $\Psi_\mathrm{o}(s,\mu,l,z)$, as
\begin{equation}
\Psi_\mathrm{o}(s,\mu,l,z)= \Omega(l,z)\, p_\mathrm{ac}(\mu,z)\, p_\lambda(l-\mu,z)\,g(\mu \, |\, s,z)\, \Phi_s(s,z) \frac{\dd V}{\dd z} \, .  \label{eq:multipsiflux}
\end{equation}
The bivariate distribution function for a flux limited sample is then
\begin{equation}
\Psi_\mathrm{o}(s,\mu)= \frac{1}{\Delta V_c} \iint \Psi_\mathrm{o}(s,\mu,l,z)\, \dd z \dd l \label{eq:psiflux}
\end{equation}
where $\Delta V_c$ is the comoving volume within the redshift range.

At this place we ignore possible redshift dependences of the individual distribution functions, postponing their discussion to section~\ref{sec:agnevo}. We also assume that the selection function is purely defined by an AGN luminosity limit and an AGN flux limit, i.e. we neglect any potential dependence of the selection function on the bulge property $s$. The selection function $\Omega$ is then given by
\begin{equation}
\Omega(l,z) = \left\{ \begin{array}{rl}
                 \displaystyle 1 & \quad \mathrm{for} \quad l \geq l_\mathrm{min} \; \& \; f \geq f_\mathrm{min} \\
                  0 & \quad \mathrm{else} 
                \end{array} \right. \ ,
\end{equation} 
where $f$ is the bolometric flux $f=l-\log(4\pi d_l^2)$ in logarithmic units.

At high redshifts and for a narrow $z$ range, this is almost identical to the luminosity limited case. However, it describes the more general case for realistic observations and remains valid when the sample covers a wider $z$ range, including the low $z$ regime. The application of this form to low redshift AGN samples is illustrated in the next subsection. Consequences for higher $z$ are discussed in section~\ref{sec:evo}.

\subsection{Application to the reverberation mapping AGN sample} \label{sec:rmsample}
We have demonstrated above that already the low $z$ AGN population will be affected by the details of the sample selection. The main difference compared to high $z$ is that for AGN in the local universe the luminosity limit is generally low and therefore the bias is less severe. Nevertheless, it is present, and as a test case we briefly estimate its influence on the AGN sample with black hole masses determined via reverberation mapping.

The technique of reverberation mapping \citep[e.g.][]{Blandford:1982,Peterson:1993} provides the most precise black hole mass estimates for type~1 AGN. It builds the foundation for the virial method through establishing a scaling relation between the size of the broad line region and the continuum luminosity \citep{Kaspi:2000,Kaspi:2005,Bentz:2009}. However, this method currently only provides measurements of $M_\bullet$ up to a scale factor depending on the (largely unknown) geometry and dynamics of the broad line region. The usual approach to determine this scale factor and thus fix the virial mass scale is to scale the reverberation mapped black hole masses of galaxies with stellar velocity dispersion measurements to the local \mbhsig relation of quiescent galaxies \citep{Onken:2004,Woo:2010}. However, this implicitly assumes that the reverberation mapping (RM) sample follows the same relation as quiescent galaxies. If its \mbhsig relation is biased by selection effects then this bias will propagate into the absolute calibration of the virial mass estimates.

\begin{figure}
\centering
\resizebox{\hsize}{!}{\includegraphics[clip]{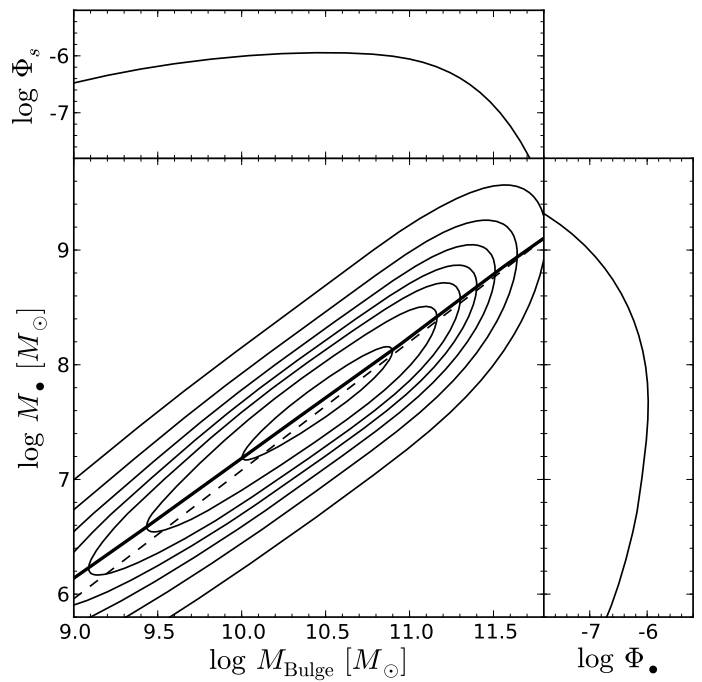}}
\caption{Predicted $M_\bullet-M_\mathrm{Bulge}$ diagram and its projections for a local type~1 AGN sample, simulating the reverberation mapping sample. A flux limit in the redshift interval ($0.003,0.15$) and a luminosity limit are applied to the sample, based on the observed range for the reverberation mapping AGN. For this sample a mild bias of 0.09~dex is estimated.}
\label{fig:bhbu_rm}
\end{figure}

\begin{figure*}
\centering
\includegraphics[width=18cm,clip]{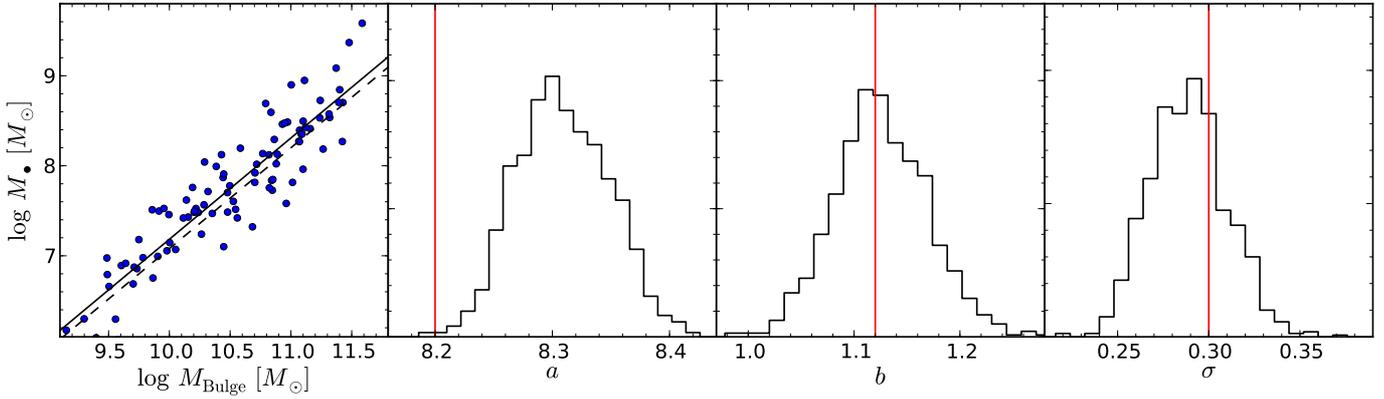}
\caption{Results of Monte Carlo simulations of the reverberation mapping AGN sample. Panels from left to right: (1) \mbhbu relation for a Monte Carlo realisation of 100 objects. The dashed line shows the input relation, the solid line the mean best fit for 1000 realisations. (2) Distribution of recovered zero points of the \mbhbu relation. The horizontal line shows the input. (3) same for the slope of the relation. (4) same for the intrinsic scatter in the relation. There is an offset in the zero point of $\sim0.1$~dex. The slope and intrinsic scatter are not strongly affected.}
\label{fig:mc_fit}
\end{figure*}

Properly defining the selection function for the RM sample is difficult, due to the heterogeneous selection of the objects. The RM sample as such is certainly not a well defined sample. Nevertheless, in order to estimate potential systematics that may be inherent in the sample, we tried to estimate the selection function, as a kind of first order approximation. The full sample covers a wide range in luminosity, from bright quasars to moderate-luminosity Seyfert galaxies. Especially for the Seyfert galaxies there is no clear AGN luminosity dependence for the selection. However, there must be an implicit limit on the AGN luminosity and also on the AGN flux. First of all, the AGN has to be luminous enough to classify the galaxy as harbouring an active nucleus. Furthermore, the continuum and broad lines have to be sufficiently bright to enable spectroscopy and the measurement of a reliable time lag. For simplicity, we relate the low luminosity limit to the faintest AGN in the sample of \citet{Bentz:2009}, NGC~4051, with $\log \lambda L_{\lambda,\mathrm{AGN}}(5100\,$\AA{}$)=41.9$. This approximately corresponds to $\log L_\mathrm{bol}=42.8$.
Equally, we use the object with the lowest flux, PG~1229$+$204, to define the flux limit. This yields a bolometric flux limit of $f_\mathrm{min}=-10.4$ in logarithmic units, which is also in good agreement with the magnitude limit of the Palomar-Green (PG) quasars in the sample \citep{Schmidt:1983,Kaspi:2000}.
We adopt a redshift range of $0.003\leq z\leq 0.15$, which contains the vast majority of the RM AGN. We also tested the effect of extending the range to $z\leq0.3$, including the full RM sample.

If we assume only a luminosity limit, the total sample bias $\langle \Delta \mu \rangle$ predicted by our model would be negligible (see Fig~\ref{fig:dmact}). Indeed, when omitting the flux limit we find $\langle \Delta \mu \rangle=0.003$, as a lower limit to the bias. However, the luminosity limit is only important for $z\lesssim0.008$. At higher $z$ the flux limit dominates the selection function. Incorporating the flux limit we get $\langle \Delta \mu \rangle=0.09$ for $z\leq 0.15$ ($\langle \Delta \mu \rangle=0.12$ for $z\leq0.3$). If we relax our conservative flux limit, the expected sample bias would decrease slightly. Decreasing the assumed luminosity limit has almost no effect, as we are dominated by the flux limit. Increasing the luminosity limit will slightly increase the bias.

In Fig.~\ref{fig:bhbu_rm} we show the predicted \mbhbu diagram. At the high mass end no bias is expected, whereas a mild bias at low masses is present. Thus, the \mbhbu relation is slightly affected.

We further tested this result with Monte Carlo simulations. We restricted our Monte Carlo sample to the same luminosity limit and flux limit and fitted the culled sample with a maximum likelihood method \citep[see e.g.][]{Gultekin:2009}, with slope $b$, normalisation $a$ and intrinsic scatter $\sigma$ as free parameters. From the restricted sample, we fitted 1000 random subsamples of 100 objects each. The distribution of the free parameters are shown in Fig.~\ref{fig:mc_fit}. We found mean values of $a=8.31$, $b=1.12$ and $\sigma=0.29$, compared to input values of $a=8.2$, $b=1.12$ and $\sigma=0.3$, i.e. we recover the predicted mean offset and confirm that the slope and scatter are not strongly affected.

Therefore, based on our model assumptions, we estimate that the reverberation mapping sample may be biased towards a high $M_\bullet/M_\mathrm{Bulge}$ ratio by $\sim0.1$~dex. This would correspond to an underestimation of the virial method by the same amount, when using the scale factor normalised to the quiescent \mbhsig relation \citep{Onken:2004,Woo:2010}. However, due to the poorly defined selection function of the RM sample this can only be taken as a qualitative evaluation of possible selection effects.

\subsection{On a black hole mass uncertainty bias} \label{sec:bhbias}
\citet{Shen:2010} discussed an additional bias that will affect AGN samples with virial mass estimates. This is a Malmquist type bias that is expected to arise from the steep turn-down of the active BHMF and the intrinsic uncertainty of virial BH mass estimates. They argue that this is independent of the luminosity bias discussed by \citet{Lauer:2007} and thus adds an additional bias to the observations. In our framework we are directly sampling from the active BHMF, thus we are already taking this effect into account. If we introduce an uncertainty in the mass estimates this will only increase the scatter, but it will not affect the mean relation $\langle \mu \rangle (s)$ or $\langle \Delta \mu \rangle$. This can again be illustrated by the Monte Carlo simulations. Adding a scatter in the BH masses does not affect our sample selection, as it is by construction based on luminosities, which can be measured with relatively high precision. Thus the uncertainty just moves the black hole masses symmetrically towards lower or higher values, without affecting the mean. 

It is also straightforward to see this from the distribution function $\Psi_\mathrm{o}(s,\mu)$. We assume that the virial mass estimate (the observed mass $\mu_\mathrm{o}$) is given by a log-normal probability distribution with mean $\mu$ and dispersion $\sigma_\mathrm{vm}$, the uncertainty in the virial mass estimate,
\begin{equation}
 g(\mu_\mathrm{o} \, |\, \mu) = \frac{1}{\sqrt{2 \pi} \sigma_\mathrm{vm}} \exp \left\lbrace - \frac{(\mu_\mathrm{o}-\mu)^2}{2 \sigma_\mathrm{vm}^2} \right\rbrace  \ . \label{eq:gvm}
\end{equation} 
The bivariate distribution function for bulge property and virial black hole mass is then
\begin{equation}
  \Psi_\mathrm{o}(s,\mu_\mathrm{o})= \int g(\mu_\mathrm{o} \, |\, \mu)\, \Psi_\mathrm{o}(s,\mu)\,\dd \mu \, . \label{eq:psivm}
\end{equation} 
The mean relation is
\begin{equation}
 \langle \mu_\mathrm{o} \rangle (s) = \frac{\iint \mu_\mathrm{o}\,g(\mu_\mathrm{o} \, |\, \mu)\, \Psi_\mathrm{o}(s,\mu)\, \dd \mu \dd \mu_\mathrm{o}}{\iint g(\mu_\mathrm{o} \, |\, \mu)\, \Psi_\mathrm{o}(s,\mu)\, \dd \mu \dd \mu_\mathrm{o}} = \frac{\int \mu\, \Psi_\mathrm{o}(s,\mu)\, \dd \mu}{\int \Psi_\mathrm{o}(s,\mu)\, \dd \mu}\ ,\label{eq:meanmvm}
\end{equation} 
when integrating over $\mu_\mathrm{o}$. This is identical to Equation~(\ref{eq:meanm}), so no additional bias is introduced by the virial mass uncertainty. 

The conceptual difference between our results and the work by \citet{Shen:2010} is, firstly, that we do not treat the black hole mass bias independently, but within the total bias budget. Secondly, we account for the fact that in observations we do not sample directly from the true BHMF, but only from the apparent BHMF affected by the relevant luminosity limits. As discussed by SW10 in detail, this directly observable distribution suffers from incompleteness at the low mass end and therefore turns over towards low $M_\bullet$.

In Fig.~\ref{fig:bhbu_vm} we show the result for a luminosity limit of $l_\mathrm{min}=47$, assuming $\sigma_\mathrm{vm}=0.3$~dex. While the black hole mass distribution is broadened, the mean relation is unchanged.

\subsection{On a spheroid uncertainty bias} \label{sec:bubias}
At least at high redshifts, not only the estimated black hole masses are significantly uncertain, but also the observed spheroid properties. For example, the emission line width of [\ion{O}{iii}] has been used several times as surrogate of $\sigma_\ast$. However, this is at best on average a reliable estimator, with a dispersion of at least $\sim 0.2$~dex \citep{Nelson:2000}. Furthermore, spheroid masses estimated for AGN host galaxies suffer from several uncertainties, including the difficulties of performing a proper AGN-host decomposition, of deblending bulge and disk components, and of converting host galaxy luminosities into stellar masses, using either a fixed mass-to-light ratio or colour information. These uncertainties may well accumulate to a total error of similar magnitude as for the black hole masses, or more.
In that case, again a Malmquist type bias is possible.

We model the observed spheroid property $s_\mathrm{o}$ by a log-normal probability distribution with mean $s$ and dispersion $\sigma_s$, analog to Equation~(\ref{eq:gvm}). The bivariate distribution function is then
\begin{equation}
  \Psi_\mathrm{o}(s_\mathrm{o},\mu)= \int g(s_\mathrm{o} \, |\, s)\, \Psi_\mathrm{o}(s,\mu)\,\dd s \, , \label{eq:psibu}
\end{equation} 
and the mean relation is 
\begin{equation}
 \langle \mu \rangle (s_\mathrm{o}) = \frac{\iint \mu\,g(s_\mathrm{o} \, |\, s)\, \Psi_\mathrm{o}(s,\mu)\, \dd s \,\dd \mu}{\iint g(s_\mathrm{o} \, |\, s)\, \Psi_\mathrm{o}(s,\mu)\, \dd s \,\dd \mu} \ .\label{eq:meanmbu}
\end{equation}
The total sample bias $\langle \Delta \mu \rangle$ is not affected by the uncertainty, as can easily be verified. However, the predicted $M_\bullet$-bulge relation is affected, as shown in Fig.~\ref{fig:bhbu_bu}. At the low mass end the mean $M_\bullet$ increases compared to the prediction without scatter, while at the high mass end it decreases. To understand this trend we have to recall that we are sampling from the luminosity-limited AGN host galaxy spheroid mass function. Due to the luminosity limit the density of sampled objects decreases at the low mass end, similarly as in the active BHMF (cf.\ SW10). Because of this decrease at the low mass end more objects are scattered from slightly larger masses to lower masses than the other way around. This produces an excess of higher true $s$ at a given observed $s_\mathrm{o}$, which in turn leads to an increase in the average $\mu$ at the given $s_\mathrm{o}$. At the high mass end the reverse happens. Due to the steep decrease towards higher masses, more black holes are scattered to higher $s_\mathrm{o}$, having on average a lower $\mu$, as they intrinsically have a lower $s$ than observed. In total, the  observed $M_\bullet$-bulge relation flattens, without changing its normalisation.

\begin{figure}
\centering
\resizebox{\hsize}{!}{\includegraphics[clip]{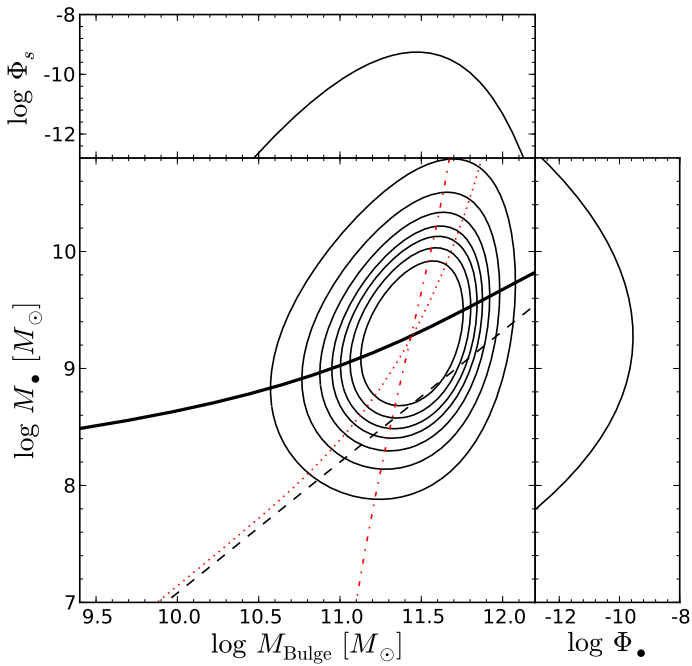}}
\caption{Predicted $M_\bullet-M_\mathrm{Bulge}$ diagram and its projections for a type~1 AGN sample, with luminosity limit of $l_\mathrm{min}=47$ (as the right panel of Fig.~\ref{fig:bhbu_agn}). For the black hole masses an uncertainty of $0.3$~dex is assumed, representing the intrinsic uncertainty in the virial method. A thick solid line shows the observed \mbhbu relation for the observed $M_\bullet$, which is identical to the relation without uncertainty in the virial mass. Only the scatter in the relation is increased. The red dashed dotted line shows the mean relation $\langle  s \rangle (\mu_\mathrm{o})$, while the dotted red line gives the mean relation $\langle  s \rangle (\mu)$, i.e. without measurement uncertainty in the black hole mass.
 }
\label{fig:bhbu_vm}
\end{figure}

For the mean spheroid mass at a given black hole mass the opposite effect happens. Uncertainties in the spheroid masses do not change the mean relation $\langle s_\mathrm{o} \rangle (\mu)$, but uncertainties in the black hole masses can strongly affect the mean relation $\langle s \rangle (\mu_\mathrm{o})$, causing a steepening of the  observed relation. This is indicated by the red lines in Figures~\ref{fig:bhbu_vm} and~\ref{fig:bhbu_bu}.

Finally, we note that in general the convolution with the uncertainty should really be done with $\Psi(s,\mu)$ rather than with $\Psi_\mathrm{o}(s,\mu)$, i.e. before applying the selection criteria. In our case of a purely luminosity limited sample, the selection function does not depend on $s_\mathrm{o}$ or $\mu_\mathrm{o}$. Therefore, our approach is justified and no additional bias is introduced for the sample in total. In case of a more complicated selection function that depends on $s_\mathrm{o}$ or $\mu_\mathrm{o}$ this might no longer be the case. An additional bias by the measurement uncertainty then becomes possible which will depend on the details of the selection function.

\begin{figure}
\centering
\resizebox{\hsize}{!}{\includegraphics[clip]{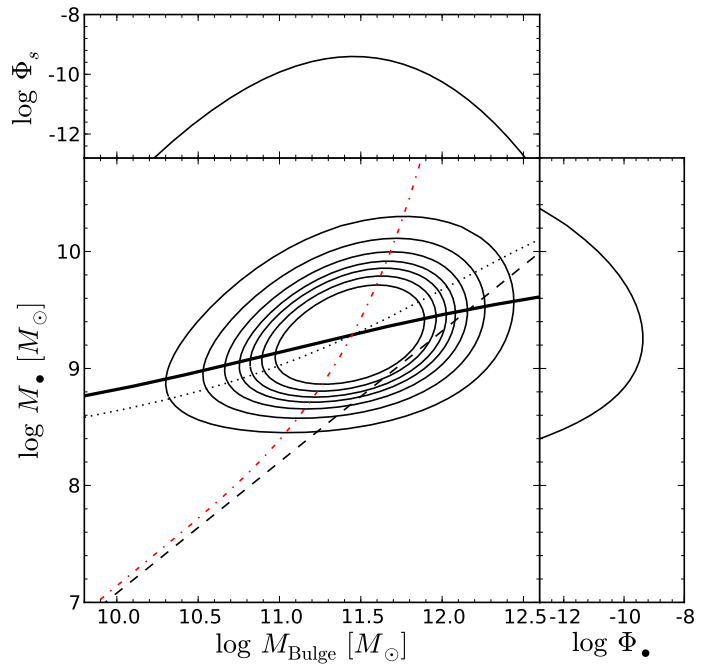}}
\caption{Same as Fig.~\ref{fig:bhbu_vm}, but with measurement uncertainty of $0.2$~dex in the spheroid mass and no uncertainty in the black hole mass. The thick solid line shows the \mbhbu relation for the observed $M_\mathrm{Bulge}$ values, while the dotted line shows the same relation but for the true $M_\mathrm{Bulge}$, i.e. for the case of no measurement uncertainty. The sample bias is not affected, but the slope of the \mbhbu relation flattens due to uncertainties in $M_\mathrm{Bulge}$. The red dashed-dotted line shows again the mean relation $\langle  s_\mathrm{o} \rangle (\mu)$, which is not affected by uncertainties in $M_\mathrm{Bulge}$.}
\label{fig:bhbu_bu}
\end{figure}

\begin{figure*}
\centering
\includegraphics[width=14cm,clip]{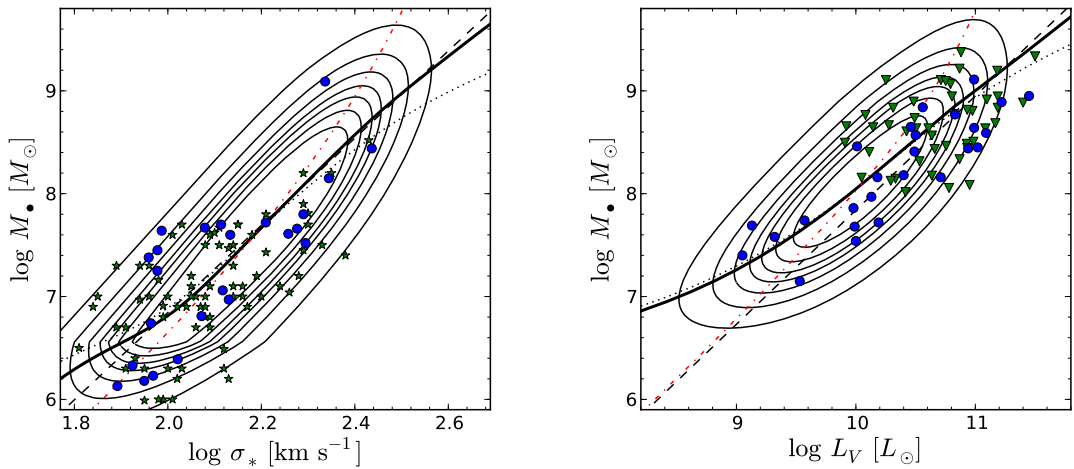} 
\caption{Comparison of our bivariate probability distribution $\Psi_\mathrm{o}(s,\mu)$ (contours), assuming reasonable numbers for the selection function, with observational data of low-redshift AGN. 
The thick black line shows the mean relation $\langle \mu \rangle (s)$, the red dashed dotted line shows $\langle s \rangle (\mu)$, the black dashed line gives the true relation. The dotted lines shows the mean relation $\langle \mu \rangle (s)$ including measurement uncertainties in $s$ of $0.05$~dex for $\sigma_\ast$ and $0.1$~dex for $L_V$. Left panel: $M_\bullet$-$\sigma_\ast$ relation. We show stellar velocity dispersion measurements for the reverberation mapping sample \citep[blue circles, ][]{Woo:2010}, and a local SDSS sample \citep[green stars,][]{Greene:2006}. Right panel: $M_\bullet-L_V$ relation for the reverberation mapping sample \citep[blue circles, ][]{Bentz:2009b} and a local QSO host galaxy sample \citep[green triangles,][]{Kim:2008}.}
\label{fig:bhbu_active}
\end{figure*}

\subsection{The slope in low-redshift AGN samples}
The black hole - bulge relations determined for low-redshift AGN samples usually plays an important role in the search for possible evolution in the relations. But they are also interesting in themselves, as it is not at all clear from a physical perspective that AGN host galaxies should follow exactly the same relation as largely inactive galaxies. Notice that the current calibration of the virial method ensures by design that AGN and non-AGN relations have no global offset from each other. But it has been noted that the slope of the $M_\bullet$-bulge relation for some local AGN samples appeared to be flatter than in the relation of black holes with dynamical black hole mass measurements \citep{Greene:2006,Kim:2008,Bentz:2009b,Woo:2010}. We were interested to see whether this flattening can be explained within our framework by selection effects, or whether an intrinsic difference in the relations between active and inactive galaxies has to be invoked.

In Fig~\ref{fig:bhbu_active} we show the AGN samples for which such evidence for a flatter slope of the $M_\bullet$-bulge relations has been diagnosed. We compare them with predictions of the bivariate distribution function $\Psi_\mathrm{o}(s,\mu)$, adopting plausible numbers for the selection function. For the \mbhsig relation we used the stellar velocity dispersion function from \citet{Sheth:2003} and assume our approximation of the selection function for the reverberation mapping sample from section~\ref{sec:rmsample}. For the $M_\bullet-L_V$ relation we adopted the galaxy luminosity function from \citet{Kochanek:2001}, converted to the V-band and into a spheroid luminosity function following \citet{Marconi:2004}. To roughly model the selection function we assume an AGN luminosity limit of $\log L_\mathrm{bol}=10^{45}$~ergs s$^{-1}$. 

Adopting these selection functions, we show  in Fig~\ref{fig:bhbu_active} the predicted mean relations $\langle \mu \rangle (s)$ as thick solid lines. The bias caused by the AGN selection depends strongly on $s$  and is largest at low $s$. Thus the relation does not only change in offset but also flattens compared to the true relation. Furthermore, measurement uncertainties in the bulge property lead to a further flattening of the slope, as discussed in \ref{sec:bubias}. Indeed, for active galaxies the bulge property is usually measured with larger uncertainty than for inactive galaxies. We computed the expected mean relations $\langle \mu \rangle (s)$ adding reasonable values for the measurement errors in the respective spheroid properties to the bivariate distributions. These are shown in Fig~\ref{fig:bhbu_active} as dotted black lines.

Considering our rough approximations to the respective sample selection functions we find a very good agreement between our prediction of a flatter $M_\bullet$-bulge relation and the observations. 
Thus we argue that these selection effects are able to reproduce the observed flattening in the slope. There is no reason to assume that AGN have generally a different black hole - bulge relation than inactive galaxies.

We note in passing that the $M_\bullet$-$L_V$ relation seems to disagree with our bivariate probability distribution at the high mass end. This may reveal a problem with our assumption, but it may also just indicate that the $M_\bullet$-$L_V$ relation is not the same for inactive galaxies and AGN host galaxies. In fact that is even expected to be so, since it is known that host galaxies of high luminosity AGN have on average younger stellar populations \citep[e.g.][]{Kauffmann:2003,Jahnke:2004,Vandenberk:2006} compared to inactive galaxies. Thus, if they obey the same $M_\bullet$-$M_\mathrm{bulge}$ relation, they actually \textit{should} deviate in the $M_\bullet$-$L_\mathrm{Bulge}$ relation.

\section{Evolution in the \mbhbu relation} \label{sec:evo}
\subsection{Evolution in a flux limited sample}
How will selection effects influence studies that test for redshift evolution in the black hole-bulge relations? If exactly the same sample selection criteria are applied at high $z$ as for a $z\approx 0$ comparison sample, no bias will be present, at least to first order (see section \ref{sec:agnevo}). Any change in the $M_\bullet$-bulge relations could then be taken as evidence for their evolution. However, this simple situation is usually not given. The local comparison is commonly provided either directly by the quiescent relation, or by a local type 1 AGN comparison sample. While the former clearly possesses different selection criteria, also the latter is not automatically selected in the same manner. This has to be taken into account when comparing low and high redshift results.

To illustrate this point we derive the sample bias for a flux limited sample at a given redshift,
\begin{equation}
 \langle \Delta \mu \rangle(z) = \frac{\iiint (\mu-a-b s) \, \Psi_\mathrm{o}(s,\mu,l,z)\, \dd l\,\dd \mu\, \dd s}{\iiint \Psi_\mathrm{o}(s,\mu,l,z)\, \dd l\, \dd \mu\, \dd s} \ , \label{eq:meandmz}
\end{equation} 
with the multivariate distribution function $\Psi_\mathrm{o}(s,\mu,l,z)$ given by Equation~(\ref{eq:multipsiflux}). We thus assume these distributions, for the moment, to be non-evolving with redshift (we discuss the more general case below). Equation~(\ref{eq:meandmz}) is equivalent to the bias for a luminosity limited sample, but with a redshift dependent luminosity limit. 

\begin{figure}
\centering
\resizebox{\hsize}{!}{\includegraphics[clip]{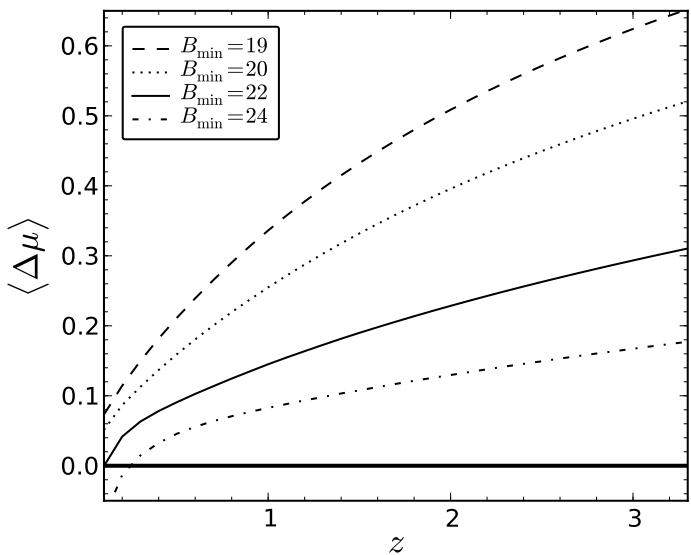}}
\caption{Redshift evolution of the selection bias for a magnitude limited AGN sample that follows the local distribution functions at all redshifts. The dashed, dotted, solid, and dashed dotted lines show $B$ band magnitude limits of $19$, $20$, $22$, and $24$~mag, respectively.}
\label{fig:dm_z}
\end{figure}

Instead of using a bolometric flux limit as before, we now adopt an optical flux limit, for example in the $B$ band. For simplicity we assume a simple power law K-correction with spectral index $\alpha=-0.44$ \citep{Vandenberk:2001}. The $B$ band luminosity is converted to bolometric luminosity using the bolometric corrections of \citet{Marconi:2004}. In Fig.~\ref{fig:dm_z} we provide the expected sample bias at a given $z$ for different apparent magnitude limits. The deeper the respective survey, the lower is the expected bias. For a fixed flux limit, the expected bias increases with redshift, which can mimic an evolutionary trend. 

Furthermore, to have a comparable bias between a local sample with, say, $B_\mathrm{min}\approx19$~mag and a sample at $z\approx2$, the high redshift sample has to be complete down to $B_\mathrm{min}\approx 24$~mag. The use of a local AGN sample as comparison sample for higher $z$ studies is already an improvement over the use of the quiescent black hole sample, but it does not ensure the absence of a selection bias. Ideally, the local comparison sample should be matched in AGN luminosity to the high $z$ sample.

On the other hand, if the selection function of the respective sample at high $z$ is known, the expected bias can be computed, and any clear offset from this prediction could be interpreted as evidence for evolution in the $M_\bullet$-bulge relations. However, we show in the following subsection that unfortunately even this approach is not fully unbiased.

\subsection{AGN evolution biases} \label{sec:agnevo}
So far we ignored any possible explicit redshift dependence of the underlying distribution functions in Equation~(\ref{eq:multipsiflux}). However, at least some of them must evolve with redshift. The stellar mass function, and thus also the spheroid mass function, are certainly changing with $z$ \citep[e.g.][]{Bundy:2005,Franceschini:2006,Pozzetti:2007,Ilbert:2010}. Furthermore, it is well known that the AGN population itself strongly evolves, apparent in the evolution of the AGN  luminosity function  \citep[e.g.][]{Ueda:2003, Hasinger:2005,Richards:2006,Bongiorno:2007,Croom:2009}. Not only the normalisation and typical luminosity change with $z$ but also the shape of the luminosity function. At low $z$ the faint end of the QSO luminosity function steepens, known as "AGN cosmic downsizing". While in the local universe the QSO luminosity function shows only mild curvature \citep{Schulze:2009}, a prominent break is present in the high $z$ luminosity function. This directly implies evolution in the active black hole mass function, the Eddington ratio distribution function, or most probably in both, leading to a change of the predicted bias with redshift, even for a fixed luminosity limit. Practically speaking it is generally not straightforward to judge if an observed evolution in the $M_\bullet$-bulge relations is caused by evolution in the intrinsic relations, or whether the trend is spurious and caused by evolution in the underlying distribution functions.

A prediction of the sample bias at given redshift requires, apart from a well defined selection function, also knowledge about these underlying distribution functions, the spheroid mass function, active BHMF and ERDF. In the local universe these distributions are at least reasonably well established (SW10), but for the high redshift universe the situation is currently less comfortable. The distribution function of velocity dispersions is reasonably well established for the local universe \citep{Sheth:2003}, but rather uncertain beyond that \citep{Chae:2010,Bezanson:2011}.
The total stellar mass function is observed up to $z\approx4$ \citep{Fontana:2006}, but galaxy mass functions for different morphological types have been determined only up to $z\approx1.4$ \citep{Bundy:2005,Franceschini:2006,Ilbert:2010}. The spheroid mass function itself is again basically unknown for higher $z$. To enable an analysis of how evolution influences the biases, we now derive very rough estimates of these distribution functions for higher redshifts. These evolving distributions are meant as illustrative only; a more detailed investigation would be beyond the scope of the current paper.

An upper limit to the spheroid mass function is given by the total stellar mass function. We use the parametric fit to the stellar mass function by \citet{Fontana:2006} for this purpose. A lower limit is given by the mass function of elliptical galaxies. To derive this distribution for arbitrary $z$ we assume the same elliptical-to-total ratio as for the local mass function from \citet{Bell:2003}, and apply this correction to the total mass function at higher redshift. This is clearly an oversimplification, as the relative contribution of elliptical galaxies seems to decrease with increasing $z$, at least to $z\sim1$ \citep[e.g.][]{Bundy:2005}. However, at higher redshifts the low mass end that is most affected by this correction is not well determined and may be underestimated if also the distribution of mass-to-light ratios changes. For the present purpose we hold that our simple approximation serves as a reasonable lower limit of the spheroid mass function. Additionally, we define a case in between these limits, using the same conversion of total mass function to spheroid mass function as assumed above for redshift zero (section~\ref{sec:loc-gen}).

There has been significant progress in the determination of the active BHMF and Eddington ratio distribution function at high redshifts in the last years \citep{Vestergaard:2008,Vestergaard:2009,Kelly:2010,Shen:2011}. But these results mainly cover the bright end of the luminosity function and thus the high mass end of the BHMF, while the low mass end is still poorly determined; also the systematics are not fully understood. Here we restrict ourselves to the use of a mass function that is consistent with current observations. 

\begin{figure*}
\centering
\includegraphics[width=18cm,clip]{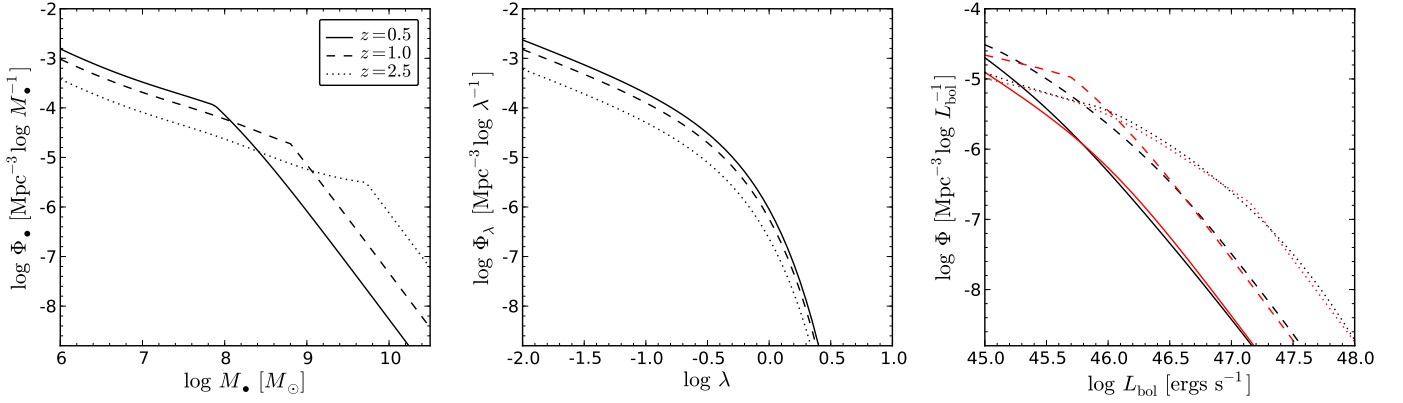}
\caption{Reconstructed AGN distribution functions at three representative redshift derived from fitting the observed AGN luminosity function to a redshift evolution model for the BHMF and ERDF. For the BHMF a mass dependent density evolution model is used, and no evolution in the ERDF is assumed. Left panel: Best fit active black hole mass function. Middle panel: Eddington ratio distribution function. The shape has been fixed, the normalisation is determined by the space density of the black hole mass function. Left panel: AGN luminosity function. The red lines show the type~1 AGN luminosity functions from \citet{Bongiorno:2007}, the black lines are our best fit to them.}
\label{fig:dfsz}
\end{figure*}

We now used the local active BHMF and ERDF as well as the redshift evolution of the type 1 AGN luminosity function (LF) as constraints to predict the distribution functions at higher redshift. To achieve a smooth redshift evolution of the BHMF and ERDF we explored several arbitrary, but reasonable evolution models for these distribution functions. We fixed the zero point of the two distribution functions to their local values (as given by SW10), and fitted their redshift evolution to the optical LF (employing Equation~\ref{eq:lf}). For the LF we used the luminosity dependent density evolution model from \citet{Bongiorno:2007}, which is based on the SDSS at the bright end and on the VVDS at the faint end. Optical $B$ band magnitudes were converted to bolometric luminosity using the bolometric corrections from \citet{Marconi:2004}. Alternatively we also explored modelling the BHMF and ERDF directly from the observed LF in individual redshift bins, which lead to the same qualitative results.

As long as only the LF is used as constraint there is clearly a degeneracy between evolution in the BHMF and in the ERDF, respectively, while presumably both are evolving. We tried to bracket this degeneracy by exploring two extreme cases. Firstly, we assumed a constant ERDF and let the BHMF change with $z$. Secondly, we fixed the active BHMF and let the ERDF evolve with $z$.

\begin{figure}
\centering
\resizebox{\hsize}{!}{\includegraphics[clip]{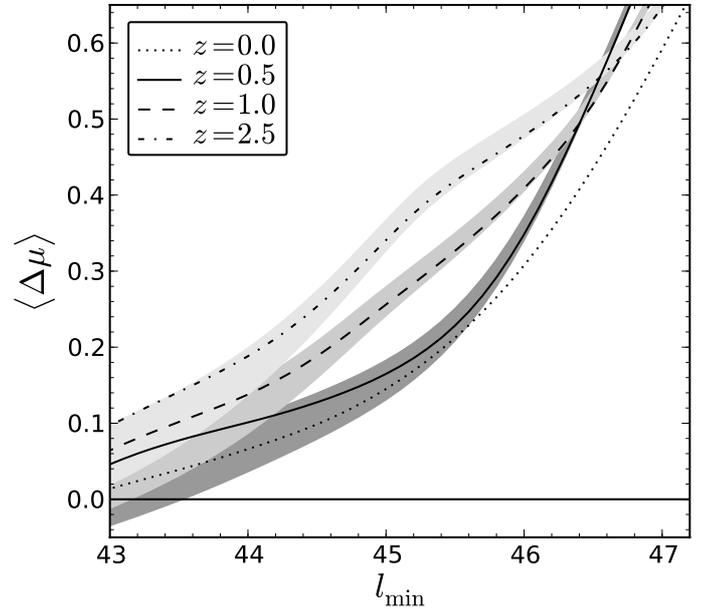} }
\caption{Expected sample bias as a function of the lower luminosity limit for the same representative redshifts as in Fig.~\ref{fig:dfsz}. The dotted line shows the $z\approx 0$ prediction, the other lines show predictions at higher $z$. The shaded area incorporates the uncertainty on the spheroid mass function, by using the total mass function as upper limit and an estimate of the elliptical mass function as lower limit.}
\label{fig:dmactz}
\end{figure}

The first case, a non-evolving ERDF, can be regarded as an upper limit to the expected AGN evolution bias. The downsizing of the AGN LF directly corresponds to a downsizing in the active BHMF. In Fig.~\ref{fig:dfsz} we show the best fit BHMFs, ERDFs, and the corresponding AGN LFs, for a few representative redshifts. Here we adopted a mass-dependent density evolution model to model the redshift dependence of the active BHMF, inspired by the LDDE model used by \citet{Bongiorno:2007} to parameterise the evolution in the AGN LF. In order to reproduce the observed AGN LF, the space density at the high-mass end has to increase substantially. Thus, the mass dependence of the active fraction is reduced, which in turn changes the active fraction bias. Because this effect tends to counteract the AGN luminosity bias, a reduced active fraction bias leads to an increase of the expected overall sample bias compared to the redshift zero case. This is illustrated in Fig.~\ref{fig:dmactz}, where we show the expected sample bias $\langle \Delta \mu \rangle$ as a function of the applied lower luminosity limit for various redshifts.

The second case, a fixed BHMF and an evolving ERDF, provides a lower limit to the expected AGN evolution bias. In this scenario the active fraction is only mildly evolving with $z$, through the evolution of the spheroid mass function. On the other hand, the ERDF needs to evolve strongly, in the sense of an increasing average $\ler$, to satisfy the constraints from the AGN LF. For a higher average $\ler$ the same luminosity limit corresponds to a lower average black hole mass limit, thus reducing the expected bias. Therefore, this assumption will predict the lowest sample bias. However, the best fit model enforces an unreasonably high space density of objects accreting at super Eddington rates and also provides a poorer fit to the AGN LF. Furthermore, it disagrees with the picture of anti-hierarchical black hole growth \citep{Marconi:2004,Merloni:2004,Merloni:2008}. Thus, this case is not physically plausible. We also experimented with intermediate evolution scenarios, where a mild evolution of the ERDF balances the increasing bias through the evolution of the BHMF to some degree. For all physically plausible scenarios, we found at least some increase of the bias compared to the local case. We conclude that the estimated sample bias for the local universe is a lower limit for high $z$ samples.

\begin{figure}
\centering
\resizebox{\hsize}{!}{\includegraphics[clip]{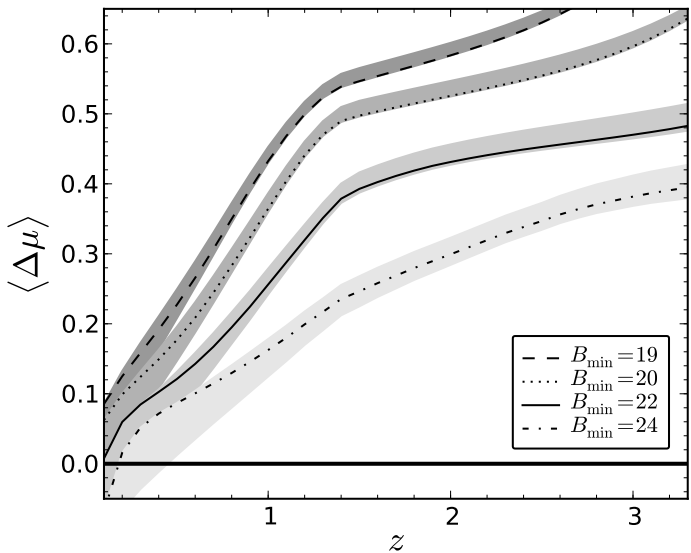}}
\caption{Redshift evolution of the selection bias for a magnitude limited AGN sample, assuming our estimate of the $z$ evolution of the underlying distribution functions, as discussed in the text. The shaded regions incorporate the uncertainty on the spheroid mass function.}
\label{fig:dm_zevo}
\end{figure}

In Fig.~\ref{fig:dm_zevo} we show the expected bias at a fixed magnitude limit as a function of redshift, equivalent to Fig.~\ref{fig:dm_z}, but now allowing for an evolving active BHMF and a constant ERDF (our case 1 above). As discussed, the resulting bias can be taken as an upper limit. Within the assumptions made in our model, only observations that show a clear excess on top of this predicted bias would constitute firm evidence for real evolution in the $M_\bullet$-bulge relations.

\begin{figure}
\centering
\resizebox{\hsize}{!}{\includegraphics[clip]{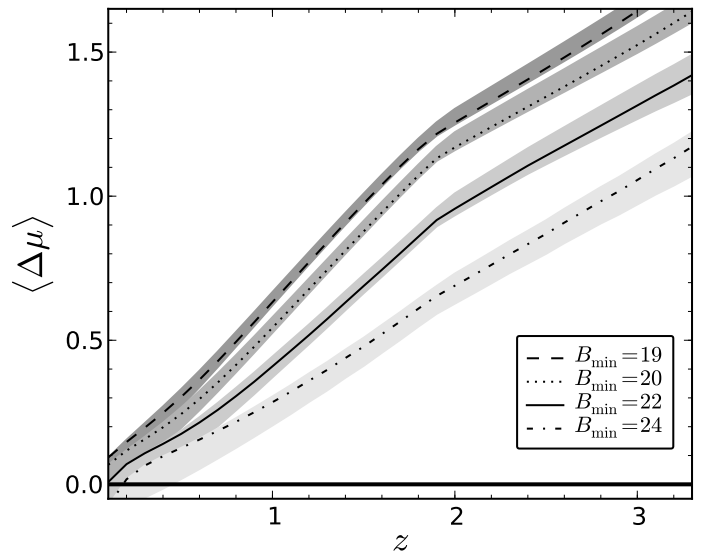}}
\caption{Same as Fig.~\ref{fig:dm_zevo}, but assuming an evolution of the intrinsic scatter in the \mbhbu relation of $\sigma(z)=0.3(1+z)^{0.5}$.}
\label{fig:dm_zevodisp}
\end{figure}

While evolution is usually expressed in terms of the normalisation of the relation, it is also possible that there is evolution in its intrinsic scatter \citep{Merloni:2010}. The strength of the sample bias is strongly affected by the amount of intrinsic scatter in the $M_\bullet$-bulge relations. Therefore, an offset of the $M_\bullet/M_\mathrm{bulge}$ ratio on top of the expected selection bias can equally be interpreted as being caused by an increased bias due to an increased intrinsic scatter, rather than as a true offset in the zero point of the respective $M_\bullet$-bulge relation. To illustrate this effect, we arbitrarily assume a redshift evolution for the intrinsic scatter as $\sigma(z)=\sigma_{z=0}(1+z)^{0.5}$ with $\sigma_{z=0}=0.3$, giving approximately $\sigma(z=2)\approx0.5$. In Fig.~\ref{fig:dm_zevodisp} we show the resulting redshift evolution of the bias for fixed magnitude limits. The predicted scatter is strongly enhanced in this case, especially at high redshifts.

An increase in the intrinsic scatter is also some kind of evolution in the relations. However, its interpretation is of course different from a change in the zero point. Whereas the latter would suggest a phase where black holes grow stronger than galaxies, or the other way around, the former would be consistent with a true coeval growth on average, in which the correlation is tightened from a rather loose one to the tight correlation we observe today. Indeed, such an evolution in scatter towards a tighter relation is expected in various models \citep[e.g.][]{Peng:2007,Volonteri:2009,Lamastra:2010}. Given sufficiently high quality data and large enough samples, it should be possible in the future to estimate the amount of evolution of the intrinsic scatter in the relation.

\section{Discussion} \label{sec:bias_discus}
Several observational studies of type~1 AGN samples found tentative evidence for an increase in the $M_\bullet/M_\mathrm{Bulge}$ ratio with redshift. This increase is often parametrised as an evolutionary behaviour $M_\bullet/M_\mathrm{Bulge}\propto(1+z)^\gamma$, with values for $\gamma$ up to $2.1$ reported \citep{McLure:2006}. As discussed in this paper, these observational studies are invariably affected by selection effects. We now apply our formalism to a few published studies. We estimate approximate distribution functions and predict the resulting sample biases. Note that we do not aim at deriving detailed corrections, or at presenting an exhaustive discussion of the relevant literature. Rather we wish to highlight the magnitudes of possible systematic biases.

Specifically, we explored two scenarios for the distribution functions that serve as upper and lower limits for the magnitude of our predicted selection bias.
\begin{itemize}
 \item[\textbullet] \textit{Model~1:} We used the local distribution functions presented in section~\ref{sec:bias_llimit} throughout the entire redshift range, i.e. we ignore any effect of redshift evolution. As discussed above, this will serve as a lower limit to the expected bias even if the scenario is unrealistic.
 \item[\textbullet] \textit{Model~2:} Here we incorporated redshift evolution in the underlying distribution functions, in particular in the AGN population. We used the model discussed in section~\ref{sec:agnevo}, assuming a mass dependent density evolution for the BHMF and a non-evolving ERDF. This model provides an approximate upper limit to the selection bias. 
\end{itemize}

\begin{figure*}
\centering
\includegraphics[width=18cm,clip]{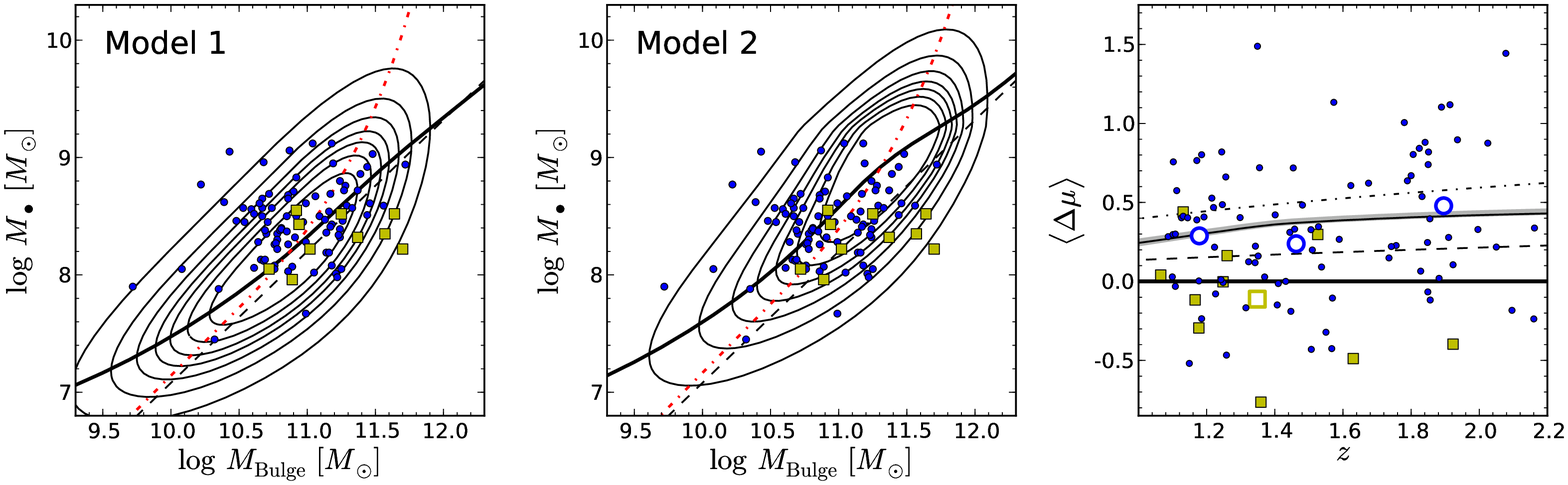} 
\caption{Left and middle panels: Predicted \mbhbu probability distributions for the samples of \citet{Merloni:2010} (blue filled circles) and \citet{Jahnke:2009} (yellow squares), using our model~1 (left) and model~2 (middle) assumptions about the distribution functions. In each panel, the black solid line shows $\langle \mu \rangle (s)$, the red dashed-dotted line shows $\langle s \rangle (\mu)$, and the thin black dashed line delineates the reference $z\approx 0$ relation. Right panel: redshift dependence of the observed \mbh/\mbu\ ratios, expressed as offset $\Delta\mu$ from the $z=0$ relation, together with the predicted evolution of the sample bias. The symbols are as before, plus the large open circles that show the mean offset for the Merloni data in three redshift bins, and the large open square showing the mean of the Jahnke data. The dashed line shows our prediction for model~1, the solid line with the shaded area shows our prediction for model~2, including the uncertainties on the spheroid mass function. The dashed-dotted line is for model~1, but assuming evolution in the \mbhbu relation $M_\bullet/M_\mathrm{Bulge}\propto(1+z)$.}
\label{fig:merloni10}
\end{figure*}

\subsection{\citet{Merloni:2010}}
A relatively large and well-defined sample was investigated by \citet{Merloni:2010}. They used an $I$ band limited set of objects drawn from zCOSMOS \citep{Lilly:2007} in the redshift range $z=[1.06,2.19]$, with  $I_\mathrm{AB}<22.5$, for which they estimated stellar masses by fitting single stellar population models to the observed multiband SEDs.  The results are shown in Fig.~\ref{fig:merloni10}. While the individual data points show considerable scatter, they are on average offset from the $z=0$ relation by $\langle \Delta \mu \rangle_\mathrm{obs}=0.34$. \citet{Merloni:2010} considered also selection biases, largely following the argument given by \citep{Lauer:2007}, as one possible origin of this offset. They concluded that their observations indicated either evolution in \mbh/\mbu\ or a significant increase of intrinsic scatter on the \mbhbu relation.

We modelled their sample selection and applied the two above mentioned models for the distribution functions to predict the expected sample bias. The correspondingly modified bivariate probability distributions are overplotted in the left and middle panels of Fig.~\ref{fig:merloni10}. Neither reproduces the distribution of data points very well, but it seems that an intermediate model might actually do a reasonable job. In the right-hand panel of the figure we compare the expected average sample offset due to selection biases, as a function of redshift, with the data points (individual and also binned into three redshift ranges). The binned mean offsets agree quite well with the range of values covered by our bias model. The prediction for a \mbh/\mbu\ ratio evolving with redshift as $\gamma = 1$, on the other hand, lies slightly above the binned data.

\begin{figure}
\centering
\includegraphics[width=7cm,clip]{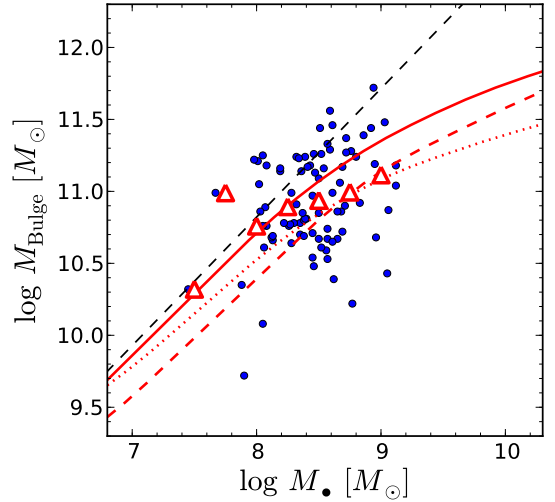} 
\caption{Comparison of predictions for the relation of mean stellar mass at given black hole mass $\langle s \rangle (\mu)$ with the data from \citet{Merloni:2010}. The blue circles show their data, the red triangles give their mean stellar masses in black hole mass bins. These are compared with predictions of $\langle s \rangle (\mu)$ using our model~2. We employ the case of no evolution in the \mbhbu relation and an intrinsic scatter of $\sigma=0.3$ (solid line), evolution in this relation with $gamma=1$ (dashed line) and for a higher intrinsic scatter of 0.5~dex but no evolution in the relation (dotted line). The black dashed line again indicates the true \mbhbu relation.
}
\label{fig:mubin_merloni10}
\end{figure}

We pointed out in Sect.~\ref{sec:bias_llimit} that the relation of mean stellar masses at given black hole mass is to first order unaffected by the AGN selection bias and depends mainly on the spheroid mass function. We binned the data points in \mbh\ accordingly and compare this binned relation  in Fig.~\ref{fig:mubin_merloni10} with the prediction for the null hypothesis (i.e., consistency with the local relation). We also computed the same predicted relations for an evolving \mbh/\mbu\ ratio (again with $\gamma = 1$), and also for a higher intrinsic scatter in \mbh/\mbu\ of 0.5~dex (but no evolution). Both curves are offset from the null hypothesis towards higher BH masses, albeit at slightly different tilts. We performed a simple $\chi^2$ test to compare the binned data with the predicted distributions. The lowest value of $\chi^2$, and in fact the only one close to a reduced $\chi^2$ of unity, was obtained for the null hypothesis. We thus conclude that this sample provides no statistically significant evidence for evolution in \mbh/\mbu\ or its scatter, and that the apparent offsets found by \citet{Merloni:2010} are consistent with the expected magnitude of AGN sample selection effects.

\subsection{\citet{Jahnke:2009}}
\citet{Jahnke:2009} studied the $M_\bullet-M_\ast$ relation for a sample of 10 AGN from COSMOS \citep{Hasinger:2007}, using optical and NIR HST images to estimate stellar masses of the host galaxies. Their sample is X-ray selected, but the spectroscopic follow-up requirement leads to a similar optical flux limit as for the work by \citet{Merloni:2010} \citep[see][]{Trump:2009}. As the covered redshift range is very similar as well, we simply added the results by \citet{Jahnke:2009} to Fig.~\ref{fig:merloni10}. \citet{Jahnke:2009} found their sample offset $\Delta\mu$ to be close to zero and concluded that their data are fully consistent with the local relation. Fig.~\ref{fig:merloni10} suggests that when accounting for selection effects, these data actually might even imply a negative evolution. It is interesting to note that also the small X-ray selected sample by \citet{Sarria:2010}, covering a similar redshift range finds no, or even a negative evolution. However, notice that both studies \citep[and also ][]{Merloni:2010} estimated only \textit{total} galaxy masses instead of \textit{bulge} masses. \citet{Jahnke:2009} discussed possible consequences of this important difference for the true evolution of the \mbhbu relation.

\begin{figure}
\centering
\resizebox{\hsize}{!}{\includegraphics[clip]{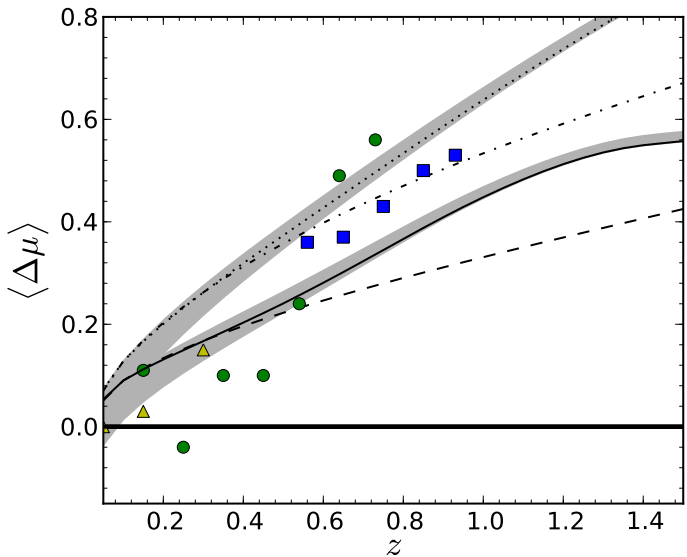} }
\caption{Observed and predicted redshift evolution of the offset in black hole masses for a sample of SDSS quasars. The green circles show the data from \citet{Salviander:2007}, using the [\ion{O}{iii}] line width as surrogate of $\sigma_\ast$, the blue squares show their results using [\ion{O}{ii}], and the yellow triangles show the results from \citet{ShenJ:2008}. The dashed and solid lines shows our model~1 and model~2, respectively, assuming no intrinsic evolution in the \mbhsig relation. The dashed-dotted and dotted lines give the predicted sample offsets for an evolution with $\gamma=1$ for the two models. The shaded areas again incorporate the uncertainty on the spheroid mass function.
}
\label{fig:salv07}
\end{figure}

\subsection{\citet{Salviander:2007}}
We now consider the study by \citet{Salviander:2007}, who studied the $M_\bullet$-$\sigma_\ast$ relation in the redshift range $0<z<1.2$, using the [\ion{O}{iii}] line width as surrogate of $\sigma_\ast$ for $z<0.8$, and the [\ion{O}{ii}] line width for $0.4<z<1.2$. Their sample is drawn from the Sloan Digital Sky Survey Data Release 3 \citep[SDSS DR3;][]{Abazajian:2005}, which does not constitute a well-defined sample \citep[see e.g.][]{Richards:2006}. Furthermore, they excluded a large fraction of objects based on quality cuts. To model their sample we simply adopted the flux limit of the SDSS main quasar sample, $i<19.1$ \citep{Richards:2002}. In Fig~\ref{fig:salv07} we show our predicted redshift evolution for the sample bias and compare them to the observations (which we augmented by similar low-$z$ measurements by \citet{ShenJ:2008}). For $z<0.6$, their data points are slightly below, for $z>0.6$ slightly above our prediction for $\Delta\mu(z)$ under the null hypothesis of no evolution in the $M_\bullet$-$\sigma_\ast$ relation. We also computed $\Delta\mu(z)$ again for the case of evolution with $\gamma = 1$.

In their discussion of possible biases in their results, \citet{Salviander:2007} attributed $\sim0.25$~dex of their offset to selection effects, which roughly agrees with our estimate at $z\sim 0.7$. They also discussed another bias caused by their signal-to-noise requirement for the narrow lines, which may accumulate to an additional $\sim0.15$~dex at high $z$. Taking this additional effect into account, the data appear again to be largely consistent with the null hypothesis of no evolution.

\begin{figure}
\centering
\resizebox{\hsize}{!}{\includegraphics[clip]{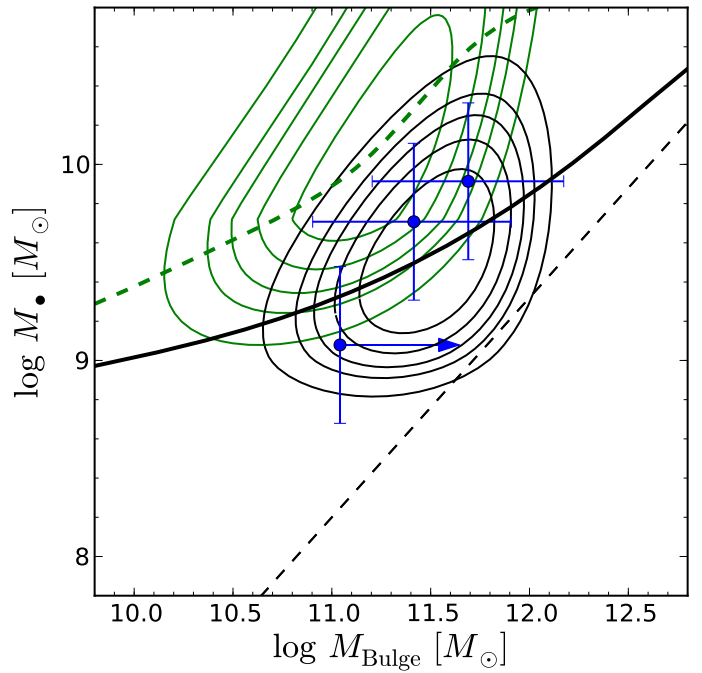} }
\caption{\mbh\ and $M_\ast$ measurements of high-luminosity quasars at $z\approx 3$ from \citet{Schramm:2008}. The black contours show the predicted \mbhbu probability distribution using model~2 for the null hypothesis of no evolution in \mbh/\mbu. For comparison, the offset green contours show the expected distribution if \mbh/\mbu were to evolve as strongly as with $\gamma = 2$.
}
\label{fig:schramm08}
\end{figure}

\subsection{High redshift luminous QSOs}
The largest apparent offsets from the local $M_\bullet$-bulge relations were so far observed for very luminous QSOs at high redshifts, $z\gtrsim3$ up to $z\sim 6$ \citep[e.g.][]{Walter:2004,Wang:2010,Targett:2011}. Because of the limitation to very luminous objects, significant selection effects are expected. At these high redshifts our knowledge of the required distribution functions becomes extremely vague, and we can do no more than guess the magnitude of the sample biases. It is nevertheless instructive to see how large they can become. We illustrate this by using the small sample of three hyperluminous QSOs around $z\sim 3$ from \citet{Schramm:2008}. These quasars originate in the Hamburg/ESO Survey \citep{Wisotzki:2000} which has an approximate magnitude limit of $B\lesssim17.5$. Stellar masses were estimated from NIR imaging that provided rest-frame $B-V$ colours.

\citet{Schramm:2008} obtained \mbh/$M_\ast$ values that were an order of magnitude larger than found in the local universe. In Fig.~\ref{fig:schramm08} we superimpose this small sample over the predicted bivariate probability distribution assuming no evolution. The data are well captured by this distribution; in fact, it is expected that luminous QSOs would nearly always be found well above the local relation. Thus, there is again no evidence even at $z\sim 3$ that the \mbhbu relation is any different from today.

However, the error bars of these and similar measurements are considerable, and some level of evolution would clearly be equally consistent with the data. To see what the expected signature for very strong evolution might be we show in Fig.~\ref{fig:schramm08} also the probability distribution for evolution with $\gamma = 2$, a value that has been suggested by some authors in the recent past. In that case, offsets of almost 2 order of magnitude would be expected; the data by \citet{Schramm:2008} are clearly inconsistent with such a scenario.

The general features discussed with this example are applicable to other studies of bright QSOs at high $z$. In particular, the $z\sim6$ SDSS main quasar sample \citep{Fan:2001} has a similar luminosity limit, and sample biases at least as strong can be expected. However, we make no attempt to guess the relevant AGN distribution functions at $z\sim 6$ and thus have to refrain from making a speculative statement about possible biases for these highest redshift objects.

\section{Conclusions} \label{sec:bias_conclu}
We investigated the ramifications of sample selection effects for the observed black hole - bulge relations. Our starting point is the bivariate probability distribution of galaxy bulge properties and black hole masses, which ideally describes the true underlying relation. However, this relation is recoverable only from a strictly volume-limited sample with completely measured galaxy properties and black hole masses. In real observations the probability distribution is inevitably modified. We incorporate the modification of the bivariate distribution by the use of an appropriate selection function which represents the observational process of sample definition and measurement. In the presence of selection effects and scatter, the true $M_\bullet$-bulge relations are generally not recovered by the observations. 

Others have recognised the importance of selection effects before, and sometimes tried to include them in their budget \citep{Salviander:2007,Merloni:2010,Bennert:2010,Lamastra:2010}. However, different papers focussed on different effects, and sometimes with quite simplifying assumptions. In this paper we go one step further and provide a common framework in which all kinds of selection effects on the $M_\bullet$-bulge relations can be investigated. We also identified some important sources of bias that were not discussed in the literature so far.

Already the widely used $M_\bullet$-bulge relations derived from observations of nearby galaxies with dynamical black hole mass measurements do not strictly reflect the underlying intrinsic relations. We showed how the bivariate probability distribution is changed due to explicit or implicit selection against galaxies where the central black hole sphere of influence is not well resolved. The resulting bias can be quite strong, almost of the order of the intrinsic scatter, when a high threshold is adopted. It becomes weaker when the threshold is lowered and more galaxies are included, but this comes at the penalty of increasing the observational errors. Measuring BH masses in galaxies with poorly resolved spheres of influence also puts higher demands on the dynamical modelling, such as the inclusion of Dark Matter halos.

For the main part of this paper we investigated in detail a number of selection effects for type~1 AGN samples, as such datasets are essential as probes for redshift evolution in the $M_\bullet$-bulge relations. We identified a variety of effects:
\begin{itemize}
\item[\textbullet] \textit{Active fraction bias:} The probability for a black hole to be in an active (type~1 AGN) state depends on the black hole mass. At least in the local universe, very high mass SMBHs are much more commonly quiescent than lower mass black holes, which have a higher probability to be actively accreting. In comparison with an ideal volume-limited set of galaxies, an AGN sample will therefore contain a higher fraction of low mass black holes. We have shown that this selection effect implies a bias towards a low $M_\bullet/M_\mathrm{Bulge}$ ratio. This source of bias has not been discussed before.

\item[\textbullet] \textit{Luminosity bias:}  The selection of AGN based on their luminosities implies a preferential selection of black holes with higher than average masses, especially when the luminosity limit is a high one. This effect has already been discussed and analysed by others \citep[e.g.][]{Salviander:2007,Lauer:2007}. Our analysis revealed that the luminosity bias depends sensitively on the details of the physically underlying distribution functions, in particular the active black hole mass function and the Eddington ratio distribution.

\item[\textbullet] \textit{Biases from measurement uncertainties:} Both black hole mass and bulge property measurements suffer from measurement and calibration errors. In combination with selection limits they may introduce additional biases.

\item[\textbullet] \textit{AGN evolution bias:} The AGN distribution functions that regulate the magnitude of the bias are evolving themselves as a function of redshift, as implied by the evolution of the AGN luminosity function. This evolution will change the expected sample bias with redshift.
\end{itemize}

These effects alter the bivariate probability distribution and can bias the conclusions drawn from observational studies. It is however important to clarify what is meant by saying that measurements of the $M_\bullet$-bulge relations are biased. These relations can be represented in a number of ways that differ in complexity. (1) The most comprehensive approach would be to investigate the full bivariate distribution of galaxy properties and black hole masses. This distribution is modified by the AGN selection, and adopting the observed relation as the measurement thus constitutes a bias with respect to the true relation. (2) The most basic approach involves computing simply the mean offset of the entire sample from the local relation. We showed that this quantity is equally biased by the AGN selection. (3) An intermediate way is to look at the projected $M_\bullet$-bulge relation, defined as the mean black hole mass for a given galaxy property. This relation also suffers from selection biases, caused by excluding black holes with certain properties from the sample during the observational process. (4) Alternatively, one can evaluate the inverse projected relation, thus the mean galaxy property at a given black hole mass. While for our parameterisation of the bivariate distribution this projection deviates from the true relation even without any selection effects, it is not modified by the AGN selection effects outlined above. Therefore this relation has the potential to be used as an unbiased estimator for the $M_\bullet$-bulge relation, as long as additional selection on the host galaxy properties do not play a role. Among these four outlined approaches, options (2) and (3) are most commonly adopted in the literature. 

It would clearly be desirable if we could properly model and correct the observations for the known sample selection effects. In order to do so, it is necessary to know the relevant underlying distribution functions, i.e. the distribution of spheroid properties such as velocity dispersions or masses; the active fraction as a function of BH mass, or alternatively the active black hole mass function; and the Eddington ratio distribution function. Thanks to recent observational progress, these distributions are now reasonably well known for the local universe; but they are still rather ill-constrained for higher redshifts.

We employed our formalism together with a set of best-guess distribution functions to investigate how much existing studies of the $M_\bullet$-bulge relation in low-redshift AGN samples might suffer from selection biases. For the heterogeneous sample of AGN with reverberation mapping data we estimated a mean offset of $\sim0.1$~dex in $M_\bullet$ at given bulge property, chiefly due to the necessary flux limitations imposed in the process of selecting these objects for observations. This is a small but not entirely negligible effect, especially in view of the important role of this sample to serve as absolute calibration of the virial method for estimating back hole masses. We also confirm that slope and scatter are not significantly affected.

Considering the more general case of low-redshift AGN samples with measured bulge properties and BH masses estimated by the virial method, we find that the observed $M_\bullet$-bulge relations suffer from a strongly mass-dependent offset from the true relation. As a consequence, the observed relation of $M_\bullet$ at given bulge property is flattened noticably. Such a flattening has been noted several times by observers \citep{Greene:2006,Kim:2008,Bentz:2009b,Woo:2010}; it can be explained in our framework without having to invoke intrinsic differences in the BH - bulge relation between active and inactive galaxies.

Extending this type of analysis to higher redshifts is hampered by our poor current knowledge of the various relevant distribution functions. In order to constrain the evolution in the $M_\bullet$-bulge relations it is crucial to not only obtain larger observed samples, but also to improve on measuring the underlying distributions. For example, not knowing the mass dependence of the active fraction at high redshifts will fundamentally limit our ability to correct for the biases associated with this effect. Furthermore, the selection function of the studied sample needs to be known; for an ill-defined heterogeneous sample the chances to properly account for the selection effects are heavily reduced.

We briefly reviewed a number of recent observational attempts to study the $M_\bullet$-bulge evolution. The result is sobering: In no case do we find statistically significant evidence for an evolving $M_\bullet$-bulge relation. While the observed apparent offsets in \mbh/\mbu\ from the local relation can be quite large, the sample selection bias estimated from our formalism is typically of the same magnitude. This does of course not exclude the possibility of real evolutionary effects that might even be already visible in some of the data; but our ability to distinguish between real and artificial trends caused by selection biases is presently not sufficient.

A possible route to circumvent several of the most problematic issues connected with AGN selection is the approach to focus on the mean galaxy property at a given black hole mass. However, while this projected distribution is formally unaffected by the AGN-specific selection biases, it also does not directly yield the intrinsic relation. In order to reconstruct the true relation in this case it is necessary to know the spheroid distribution function and the intrinsic scatter in the $M_\bullet$-bulge relation. Neither is well constrained at high redshifts, but these quantities are observables, so that this approach may become very powerful in the future.

\begin{acknowledgements}
We thank Chien Peng for fruitful discussions on this manuscript. 
We acknowledge support by the Deutsche Forschungsgemeinschaft under its priority programme SPP1177, grant Wi~1369/23. 
\end{acknowledgements}


\begin{appendix}

\section{Validation of the bivariate probability distribution}
\begin{figure*}
\centering
\includegraphics[width=18cm,clip]{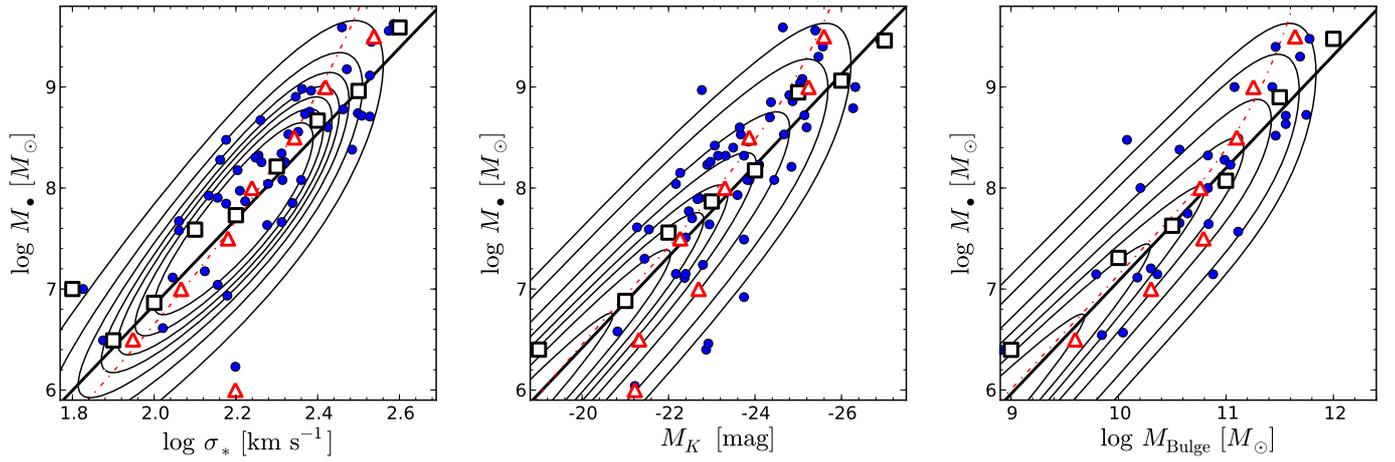} 
\caption{Comparison of our bivariate probability distribution $\Psi(s,\mu)$ (contours) with observational data of  black holes with dynamical mass measurements (blue circles). The thick black line shows the mean relation $\langle \mu \rangle (s)$, the red dashed dotted line shows $\langle s \rangle (\mu)$. The black open squares gives the mean $\mu$ binned in $s$ from the observations (an estimate of $\langle \mu \rangle (s)$) and the red open triangles gives $s$ binned in $\mu$ (an estimate of $\langle s \rangle (\mu)$). Left panel: $M_\bullet-\sigma_\ast$ sample from \citet{Gultekin:2009}. Middle panel: $M_\bullet-M_K$ sample from \citet{Hu:2009}. Right panel: $M_\bullet-M_\mathrm{bulge}$ sample from \citet{Haering:2004}. }
\label{fig:bhbu_quiet}
\end{figure*}

Our chosen parameterisation of the bivariate probability distribution $\Psi(s,\mu)$, given as
\begin{equation}
 \Psi(s,\mu)=g(\mu \, |\, s)\, \Phi_s(s) \ ,  \label{eq:psi-app}
\end{equation} 
(cf.\ Equation~(\ref{eq:psi})), follows common practice, but it is not the only possible solution. In particular, this parameterisation produces the visual impression of an upturn in the contours towards high values of $s$, corresponding to an upwards curved ridge line of the distribution. This upturn is simply a consequence of the fact that the space density of galaxies is steeply declining towards high masses or velocity dispersions. Since the quantitative conclusions in this paper depend on the appropriateness of the adopted form for $\Psi(s,\mu)$, it is worthwhile to test if our parameterisation is actually consistent with current observations. Here we perform such a test.

In Fig.~\ref{fig:bhbu_quiet} we superimpose our intrinsic bivariate probability distribution $\Psi(s,\mu)$, which includes an estimate of the relevant bulge property distribution function, with data of galaxies for which dynamical black hole mass measurements are available. We consider the distribution of $M_\bullet$ versus the three most commonly measured galaxy properties: Stellar velocity dispersions $\sigma_\ast$, $K$ band absolute magnitudes $M_K$, and stellar bulge masses $M_\mathrm{bulge}$. For the $M_\bullet$-$\sigma_\ast$ relation we incorporated the stellar velocity dispersion distribution function from \citet{Sheth:2003} and compare the result to the observational sample from \citet{Gultekin:2009}. For the $M_\bullet$-$M_K$ relation we employed the 2MASS K-band luminosity function from \citet{Kochanek:2001} and show the sample from \citet{Hu:2009} for comparison, and for the $M_\bullet$-$M_\mathrm{bulge}$ relation we use our estimate for the spheroid mass function, based on galaxy mass functions from \citet{Bell:2003}, and compare this with the sample from \citet{Haering:2004}. 

It is important to note that in this comparison we intentionally do not consider any sample selection effects. Thus, the distribution of data points in the three diagrams does not, and cannot, agree with the overall probability distribution. The test that we want to perform focuses on the upper right corner of each diagram, and the question is: Do the data follow the mentioned upturn in the ridge line? While there is of course considerable observational and intrinsic scatter in the data, a visual inspection of Fig.~\ref{fig:bhbu_quiet} clearly confirms that indeed there is such a trend. 

We can perform a more quantitative test of the adopted ansatz by separately looking at the two univariate projections of the bivariate distribution, $\langle \mu \rangle (s)$ (Equation~(\ref{eq:meanm}) -- represented by the thick solid lines in Fig.~\ref{fig:bhbu_quiet}), and $\langle s \rangle (\mu)$ (Equation~(\ref{eq:means}) -- shown by the red dashed-dotted lines. To compare these curves with the measurements we computed the mean $\mu$ in bins of $s$ and the mean $s$ in bins of $\mu$ from the data, shown in Fig.~\ref{fig:bhbu_quiet} as black open squares and red open triangles, respectively. Towards high values of $s$ there is a  clear trend for the triangles to be located above and to the left of the squares, and the predicted curves trace the points quite well. We performed a simple $\chi^2$ test on all points above a certain minimum value of $s$ or $\mu$, respectively, and found that the binned data are consistently described by the predicted curves, with a reduced $\chi^2$ always of order unity independently of the chosen cutoff. We conclude that our adopted form for the bivariate probability distribution $\Psi(s,\mu)$ is consistent with the current observations.
\end{appendix}

\end{document}